\PassOptionsToPackage{prologue,dvipsnames}{xcolor}
\documentclass[sigconf]{acmart}
\acmConference[ISSTA 2024]{ACM SIGSOFT International Symposium on Software Testing and Analysis}{16-20 September, 2024}{Vienna, Austria}
 
\definecolor{pitchblack}{cmyk}{1 1 1 1}

\AtBeginDocument{%
  \providecommand\BibTeX{{%
    \normalfont B\kern-0.5em{\scshape i\kern-0.25em b}\kern-0.8em\TeX}}}
\AtBeginDocument{\color{pitchblack}} 
\usepackage[multiple]{footmisc}
\usepackage{tabularx}
\usepackage{multirow}
\usepackage[dvipsnames]{xcolor}
\usepackage{color}
\usepackage{tcolorbox}
\usepackage{makecell}
\usepackage{listings,lstautogobble}
\usepackage [english]{babel}
\usepackage [autostyle, english = american]{csquotes}
\MakeOuterQuote{"}
\usepackage{enumitem}
\setlist{leftmargin=5.5mm}
\usepackage{nameref}
\usepackage{marginnote}
\usepackage{enumitem}

\usepackage{pifont}% http://ctan.org/pkg/pifont

\usepackage{titlesec}
\titlespacing*{\section}{0pt}{0.6\baselineskip}{0.2\baselineskip}
\titleformat{\section}
  {\normalfont\Large\bfseries} 
  {\thesection}{1em}{\MakeUppercase} % \MakeUppercase makes the title uppercase

\usepackage{multicol}

\newcommand{\cmark}{\ding{51}}%
\newcommand{\xmark}{\ding{55}}%

\newcommand{\SATFull}[1]{static application security testing tool#1}
\newcommand{\SATFullC}[1]{Static application security testing tool#1}
\newcommand{\SAT}[1]{SAST#1}

\newcommand{\RQOneTopic}{Effectiveness}
\newcommand{\RQTwoTopic}{Warning-based Prioritization}
\newcommand{\RQThreeTopic}{Waiting Time}

\newcommand{\RQOne} {How effectively can \SAT{s} detect vulnerabilities in vulnerability contributing commits?}
\newcommand{\RQTwo} {How can we use \SAT{} warnings to prioritize changes for secure code reviews?}
\newcommand{\RQThree} {How much computation time is required?}
\newcommand{\smallheading}[1]{\vspace{2mm}\noindent\textbf{#1} \\}
\newcommand{\inlineheading}[1]{\vspace{2mm}\noindent\textbf{#1}}
\newcommand{\inlineheadingNoSpace}[1]{\noindent\textbf{#1}}
\newcommand{\textbox}[1]{
    \begin{tcolorbox}[colback=gray!5!white,colframe=gray!100!black]
        #1
    \end{tcolorbox}
}
\newcommand{\goodNumbers}[1]{\textcolor{OliveGreen}{#1}}
\newcommand{\badNumbers}[1]{\textcolor{Maroon}{#1}}

\newcommand{\reviewersModification}[1]{\textcolor{pitchblack}{#1}}

\newcommand{\reviewersMinorModification}[1]{\textcolor{pitchblack}{#1}}

\begin{document}

\title[An Empirical Study of Static Analysis Tools for Secure Code Review]{An Empirical Study of Static Analysis Tools for \\
Secure Code Review}

% \title{
% % Do Tools Help The Reviewers Who Cry Wolf:
% An Empirical Study of Static Analysis Tools for \\
% Secure Code Review
% Supporting Vulnerability Detection in Code Changes
% In-Progress 02:
% Static Analysis Tool in Practice: Code Review > Manual Effort > Security Issues
% Comparison and Evaluation on Software Analysis Tools for Secure Code Reviews
% Exploring Automated Analysis Tools for Code Review Enhancement
% Supporting Modern Code Review with Automated Analysis Tools
% Do static analysis help developers identify security issues during code reviews?
% Static Analysis in Code Review Practices: Will It Help Reviewers Find Security Issues?
% Evaluating Static Analysis for Code Reviews: Will It Help Developers Find Security Issues?
% Comparison of Static Analysis Tools for Helping Developers Find Vulnerabilities in Code Changes
% }

% \author{
%     \IEEEauthorblockN{
%         Author1\IEEEauthorrefmark{1}, 
%         Author2\IEEEauthorrefmark{2}, 
%         Author3\IEEEauthorrefmark{2}, 
%         Author4\IEEEauthorrefmark{2}
%     }
%     \IEEEauthorblockA{
%         School of Computing and Information Systems, University of Melbourne
%         \\\{1\}@student.unimelb.edu.au\IEEEauthorrefmark{1}
%         \\\{2, 3, 4\}@unimelb.edu.au\IEEEauthorrefmark{2}
%     }
% }

\author{Wachiraphan Charoenwet}
\email{wcharoenwet@student.unimelb.edu.au}
\affiliation{%
  \institution{The University of Melbourne} \country{Australia}
}
\author{Patanamon Thongtanunam}
\email{patanamon.t@unimelb.edu.au}
\affiliation{%
  \institution{The University of Melbourne}
  \country{Australia}
}
\author{Van-Thuan Pham}
\email{thuan.pham@unimelb.edu.au}
\affiliation{%
  \institution{The University of Melbourne}
  \country{Australia}
}
\author{Christoph Treude}
\email{ctreude@smu.edu.sg}
\affiliation{%
  \institution{Singapore Management University}
  \country{Singapore}
}

\begin{abstract}

% ORIGINAL
% Identify area
% Identifying security issues early for software development would reduce latent impacts on software systems.
% Code review is a manual code analysis approach used by numerous software projects to identify various issues.
% Previous works argue that automated \SATFull{s} (\SAT{s}) may help reviewers detect more security issues. 
% % Various \SAT{s} have been evaluated to understand their vulnerability detection capability.
% % the large amount of warnings produced by \SAT{s} concerns the users. 
% % Identify research gap
% However, little work has yet investigated the practical capacity of \SAT{s} to support secure code review.
% % Additionally, prior works usually evaluated \SAT{s} with vulnerable versions of the subject programs that may accumulate the code changes from multiple commits.
% % Since \SAT{} warnings are usually inspected by developers, the required effort to use tools is one of the key concerns that may hinder the adoption of \SAT{s} by developers.
% Furthermore, most \SAT{} studies are also based on synthetic programs or fully vulnerable versions of subject programs which are not representative of real-world code changes in the code review process. 
% % and can potentially yield results that favor some tools that implement certain techniques.

% REVISE
Early identification of security issues in software development is vital to minimize their unanticipated impacts. 
Code review is a widely used manual analysis method that aims to uncover security issues along with other coding issues in software projects. 
While some studies suggest that automated \SATFull{s} (\SAT{s}) could enhance security issue identification, there is limited understanding of \SAT{}'s practical effectiveness in supporting secure code review. 
Moreover, most \SAT{} studies rely on synthetic or fully vulnerable versions of the subject program, which may not accurately represent real-world code changes in the code review process.

% Outcome / Novelty / Contribution
% ORIGINAL
% To address this research gap, we analyze C and C++ \SAT{s} with a novel dataset from actual code changes that contribute to exploitable vulnerabilities. 
% In addition to the effectiveness of \SAT{s}, we quantified the performance of code review to identify security issues when \SAT{} warnings are available.
% % To develop the dataset, we systematically annotate 5,354 commits from 341 C and C++ projects in the Vulnerability Contributing Commits (VCC) dataset. 
% In total, the selected dataset consists of 319 real-world vulnerabilities with 815 vulnerability-contributing commits (VCCs) in 92 C and C++ projects. 
% Using the developed dataset, we analyze 5 \SAT{s}.
% Our results show that \textcolor{red}{a single \SAT{} can produce warnings in vulnerable functions of 52\% of VCCs.
% % Combining \SAT{s} can increase the detected VCCs up to 71\% to 77\%. 
% More importantly, we find that warning-based prioritization for code reviews can improve the code review performance.
% However, 22\% of VCCs are still undetected even when multiple tools are used because of limitations in \SAT{} checking rules.
% % especially VCCs related to vulnerabilities of type protection mechanisms, access control, and control flow management.
% Finally, the
% % can be incorporated into automated testing when the code is submitted because its 
% average computing time of \SAT{s} in our experiment is less than 45 minutes.
% We discuss our findings and provide suggestions for reviewers, researchers, and \SAT{} developers.
% }

% REVISE
To address this gap, we study C/C++ \SAT{s} using a dataset of actual code changes that contributed to exploitable vulnerabilities. 
Beyond \SAT{}'s effectiveness, we quantify potential benefits when changed functions are prioritized by \SAT{} warnings.
Our dataset comprises 319 real-world vulnerabilities from 815 vulnerability-contributing commits (VCCs) in 92 C and C++ projects. 
The result reveals that a single \SAT{} can produce warnings in vulnerable functions of 52\% of VCCs. 
Prioritizing changed functions with \SAT{} warnings can improve accuracy (i.e., 12\% of precision and 5.6\% of recall) 
and reduce Initial False Alarm (lines of code in non-vulnerable functions inspected until the first vulnerable function) by 13\%.
% allow reviewers to reach vulnerable changes faster in a limited reviewing effort. 
\reviewersModification{Nevertheless, at least 76\% of the warnings in vulnerable functions are irrelevant to the VCCs, and 22\% of VCCs remain undetected due to limitations of \SAT{} rules.}
% The average \SAT{} computing time is less than 45 minutes. 
Our findings highlight the benefits and the remaining gaps of \SAT{}-supported secure code reviews and challenges that should be addressed in future work.

\end{abstract}

\begin{CCSXML}
<ccs2012>
   <concept>
       <concept_id>10002978.10003022.10003023</concept_id>
       <concept_desc>Security and privacy~Software security engineering</concept_desc>
       <concept_significance>500</concept_significance>
       </concept>
   <concept>
       <concept_id>10011007.10011074</concept_id>
       <concept_desc>Software and its engineering~Software creation and management</concept_desc>
       <concept_significance>500</concept_significance>
       </concept>
 </ccs2012>
\end{CCSXML}

\ccsdesc[500]{Security and privacy~Software security engineering}
\ccsdesc[500]{Software and its engineering~Software creation and management}

\keywords{Code Review, Static Application Security Testing Tool, Code Changes Prioritization}

\maketitle

\section{Introduction}
\label{Introduction}

% KEY POINTS
% 1) “The biggest challenge of the review process is the manual effort required from reviewers”  [Singh et al.]
% 2) SAT can help identify issues [Panichella et al.] but it may also demand extra computation time & manual effort 
%   Still, security issues are not highlighted in the code review context
% Most existing SAT evaluations already focus on effectiveness, yet to consider other costs (though some may have indirectly considered a few costs e.g., execution time)
%   In addition, the existing benchmarks are isolated from the development context (mostly used specific versions of programs as subjects, not the commits)
% lack of prior studies on how static analysis and dynamic analysis can enhance code review \cite{Iannone2022TheSoftware}

% Identifying security issues in software is one of software quality assurance's goals. 
Managing security issues in software products is crucial because such issues, especially exploitable vulnerabilities, can exponentially impact end users and require more resources to resolve if discovered in a later stage.
Aligned with the Shift-Left concept, which advocates for the early issue detection in software~\cite{Weir2022ExploringResponsibility}, previous work~\cite{Braz2022SoftwarePerspective} argues that manual code review is an approach adopted by numerous software projects to identify and mitigate issues before merging new code changes into the existing codebase. 
% Reviewers can freely inspect and provide comments on the code changes authored by developers.
Unlike traditional practice, modern code review lacks concrete requirements~\cite{Beller2014ModernFix} and can take multiple iterations to complete. 
The effort that reviewers, who typically have many resource constraints, must invest poses a persistent challenge for secure code reviews (i.e., finding security issues in the reviewed code). 
Indeed, identifying issues that could contribute to vulnerabilities often requires security knowledge and meticulous inspection from reviewers~\cite{Braz2022SoftwarePerspective}, and hence fortifies the challenges.

% Previous studies \cite{Goseva-Popstojanova2015OnVulnerabilities, Harman2018FromAnalysis, Ernst2004StaticDuality, Sadowski2015Tricorder:Ecosystem, Aloraini2019AnTools} suggests that \SAT{s} can detect security-related issues and exploitable vulnerabilities. 
Although prior code review studies \cite{Singh2017EvaluatingEffort, Mehrpour2022CanDefects, Panichella2015WouldReviews} suggested that 
% \SATFull{s} (\textit{\SAT{s}}, hereafter) 
static analysis tools
could help reviewers identify issues related to coding styles, none have explored their effectiveness 
% in addressing security aspects.
as the \SATFull{s} (\textit{\SAT{s}}, hereafter). 
Conversely, current \SAT{} studies~\cite{Lipp2022AnDetection, Stefanovic2020StaticReview, Aloraini2019AnTools, Nunes2017OnStudy, Aloraini2017EvaluatingApps} have investigated \SAT{}'s effectiveness based on the released software versions that contain exploitable vulnerabilities.
The existing works have not examined the tools with vulnerabilities that gradually appear during the software development process.
% However, they only measure the effectiveness of \SAT{s} given partially synthetic source code or realistic source code from a released version of software systems that contain exploitable vulnerabilities.
% \textcolor{red}{In the code review context, code changes can be small and tend to contain security issues that, although contribute to vulnerabilities, are not yet fully exploitable.}
In code reviews, changes are typically small~\cite{Rigby2013ConvergentPractices} containing security issues that
% , although contribute to the bigger vulnerabilities, 
are both difficult to identify and easy to overlook~\cite{Braz2022SoftwarePerspective}.
Moreover, a complete vulnerability may require the contribution of code changes from multiple code commits~\cite{Iannone2022TheStudy}.
With the different contexts and nature of code reviews,
% compared to previous studies
it remains unclear to what extent \SAT{s} can cost-effectively assist reviewers in finding security issues.
% in code changes.
\reviewersModification{
To the best of our knowledge, prior works have not investigated the effectiveness of the \SAT{s} using code commits, and for code reviews.
}

% Indeed, the realistic datasets are crucial for the \SAT{} evaluation that incorporates the code review aspects. 
% A recent study~\cite{Lipp2022AnDetection, Li2023ComparisonJava} highlights that existing benchmark datasets for \SAT{} lack the complexity of the source code because they are mostly synthetic or partially synthetic. 
% The realistic benchmark derived from the diverse real-world programs can also improve the generalization of benchmarking results~\cite{Metzman2021FuzzBench:Service}.
% Additionally, little is known about the extent to which the \SAT{s} can detect the issues in the early code changes, which are the targets of code reviews, that can introduce potential security issues. 
% The \SAT{} studies~\cite{Lipp2022AnDetection, Thung2012ToTools, DabruzzoPereira2020OnProjects} only evaluated \SAT{s} with real-world vulnerabilities in certain versions of the systems that are the result of multiple code changes over time.

% what we do
We investigate the capability of \SAT{s} to support secure code reviews by addressing three key aspects: \textcircled{1} effectiveness of \SAT{s} on vulnerable code commits, \textcircled{2} potential benefits of \SAT{}-supported code reviews,
and \textcircled{3} associated waiting time.
% We select code commits that contributed to 1,768 exploitable vulnerabilities of 338 C and C++ projects from the Vulnerability Contributing Commits (VCC) dataset~\cite{Iannone2022TheStudy}. 
% We systematically annotate the vulnerable changes in \textcolor{red}{312,258} changed functions in \textcolor{red}{59,138} changed files of \textcolor{red}{5,207} commits.
Our empirical study uses 815 real-world \textit{vulnerability-contributing commits} (VCCs) in 92 C and C++ software projects.
These VCCs involve 1,060 vulnerable functions 
% in \textcolor{red}{9,855} changed files
which contributed to 319 exploitable vulnerabilities.
With this dataset, we study five commonly used C and C++ \SAT{s}~\cite{Lipp2022AnDetection}, which employ diverse static analysis techniques to detect vulnerabilities, i.e., Cppcheck, CodeChecker, CodeQL, Flawfinder, and Infer. 
To measure the effectiveness of \SAT{s}, we analyze how many VCCs the tools can warn about.
\reviewersModification{For the potential benefits of code reviews supported by \SAT{s}, we analyze how many vulnerable functions can be identified within a limited code review effort when changed functions are prioritized based on the information in the \SAT{} warnings~\cite{Hong2022WhereTo} (henceforth, \textit{warning-based prioritization)}.}
To do so, we measure the accuracy using precision and recall at a fixed review effort (i.e., 25\% of lines of code in changed functions) and Initial False Alarm (IFA)---percentage of lines of code in warned functions until the first vulnerable functions.
Additionally, we measure the time that each tool takes to analyze the VCCs.
% The results show that a single \SAT{} may help reviewers identify vulnerability in more than half of the VCCs because Flawfinder can produce warnings in 52\% of the vulnerable functions and 89\% of the vulnerable files across 815 VCCs.

\reviewersModification{
Our results show that a single \SAT{} can produce warnings in the vulnerable changes (i.e., changed files and changed functions) of more than half of the VCCs. Specifically, Flawfinder can produce warnings in at least one vulnerable function for 52\%, and one vulnerable file for 89\%, of 815 VCCs.
}
Combining warnings from multiple \SAT{s} can achieve a higher vulnerability detection rate, i.e., 78\% at the function level.
% Particularly, VCCs related to vulnerabilities of type access control and control flow management are rarely detected.
% \textcolor{red}{Although some studied \SAT{s} should detect the vulnerability, none of the VCCs related to the type protection mechanism failure can be detected because the \SAT{} rules do not fit the implementations of protection mechanisms in particular projects}.

With respect to code reviews, prioritizing changed functions using the warnings from \SAT{s} during code reviews can enhance the discovery of vulnerable changes by increasing up to 12\% of precision and 5.6\% of recall and reducing up to 13\% of IFA at 25\% of code review effort.
Lastly, the computation time of \SAT{s} should fit within the code review waiting period since the average computation time on the selected VCCs ranges from 20 seconds to 45 minutes, with a notable increase when the size of projects reaches 50k-100k LOC.
\reviewersModification{Nevertheless, \SAT{s} still need an improvement because at least 76\% of the warnings are irrelevant to the vulnerability in the corresponding VCCs.
Additionally, 22\% of VCCs (10\% of the exploitable vulnerabilities) do not receive any warnings from any of the five studied tools.}
% This indicates that \SAT{s} can be integrated into the automated tests (i.e., CI/CD pipeline) to support code reviews.

\inlineheading{Novelty and Contribution:}
% A real-world vulnerability dataset (xxx vulnerabilities from xxx project) for benchmarking \automatedTool{s}~ 
% An effort-sensitive evaluation approach for \automatedTool{s}
To the best of our knowledge, this paper is the first to:
\begin{itemize}[noitemsep,topsep=1mm]
    \item \reviewersModification{Investigate the effectiveness of C and C++ SASTs on the code commits, which are subject to code reviews, that introduced real-world vulnerabilities in 92 projects of diverse application domains, covering eight types of CWE vulnerabilities}
    \item \reviewersModification{Investigate the potential benefits of warning-based-prioritization for vulnerability identification in VCCs' changed functions during code reviews}
    % \item Analyze the computing time of C and C++ \SAT{s} on the 92 real-world projects of diverse application domains
    \item \reviewersModification{Provide a versatile and extensible automated benchmarking framework for \SAT{} evaluation using open-source code commit datasets in various programming languages~\cite{Charoenwet2024AnCommits}}
    \item Release an annotated dataset of 1,064 vulnerable files and 1,060 vulnerable functions in 815 commits~\cite{Charoenwet2024DatasetReview}

\end{itemize}

% MERGED DEFINITION WITH BACKGROUND
% \inlineheading{Definitions:}
% We briefly describe the meaning of the frequently used terms in this paper to prevent ambiguity:
% \begin{itemize}
%     \item \textbf{\SATFullC{} (\SAT{})}: A type of quality assurance tool that can identify certain issues in source code without executing the program. We consider \SAT{s} that employ two analysis techniques: \textcircled{1} syntactic-based and \textcircled{2} semantic-based ~\cite{Lipp2022AnDetection}.
%     \item \textbf{Vulnerability contributing commit (VCC)}: A set of code changes comprises the modified lines of code that contributed to Common Vulnerabilities and Exposures (CVE) vulnerability. These changes are stored in a code commit that is linked to a version of the subject program. One CVE vulnerability can be introduced by multiple VCCs.
% \end{itemize}

\inlineheading{Paper Organization:} 
The rest of this paper is organized as follows: Section \ref{background} explains the background of this study. 
Section \ref{study_design} presents our study design. 
% Section \ref{dataset_preparation} explains the data preparation. 
% Section \ref{analysis} explains the analysis process for the warning-based prioritization code review. 
Section \ref{results} presents our study approaches and results of each research question. 
Section \ref{discussion} discusses the findings. 
Section \ref{threats} reports threats to validity. 
Section \ref{related_work} discusses related work. 
Finally, Section \ref{conclusion} draws the conclusion.

\section{Background and Definitions}
\label{background}

% We briefly describe the challenges of the secure code review process and the roles of \SAT{s} in code reviews in this section.

% \textcolor{red}{TODO - Restructure to align with introduction and clearly highlight gaps/differences of this work}
% Key term: Code changes

% Security in development cycle - Shift left concept
% Code reviews -> find issues in code changes, aligned with shift-left
% But there are challenges in secure code review

% Studies have suggested tool support
% SAT complements issue identification at the code changes level, but not the from security aspect
% Studies that look at the security aspect of SAT not look at code changes
% We're filling this gap (not showing our results -> our results can be shown in related works)

   % \item \textbf{\SATFullC{} (\SAT{})}: A type of quality assurance tool that can identify certain issues in source code without executing the program. We consider \SAT{s} that employ two analysis techniques: \textcircled{1} syntactic-based and \textcircled{2} semantic-based ~\cite{Lipp2022AnDetection}.
   %  \item \textbf{Vulnerability contributing commit (VCC)}: A set of code changes comprises the modified lines of code that contributed to Common Vulnerabilities and Exposures (CVE) vulnerability. These changes are stored in a code commit that is linked to a version of the subject program. One CVE vulnerability can be introduced by multiple VCCs.
    
In this section, we briefly describe the key concepts and terms used in this paper and discuss the limitations and challenges.

\inlineheading{Secure code review} refers to
% Security issues in software systems can cause more harm if discovered late.
a code review practice that focuses on security aspects of the software system.
This practice conforms to the shift-left concept that promotes early quality assurance in software development~\cite{Weir2022ExploringResponsibility} to reduce the impacts of undiscovered security issues.
% Manual code review is an approach that many projects have adopted~\cite{Bacchelli2013ExpectationsReview} for this purpose.
% Code review is a common software quality practice that aligns well with the shift-left concept.
During the development cycle, developers submit a code commit that contains small code changes 
% (25-263 lines on average
\cite{Rigby2013ConvergentPractices}
% ) 
to be reviewed by other developers before it can be integrated into the codebase. 
Reviewers can inspect the code changes and provide feedback on various aspects, including security~\cite{Thompson2017AProjects}, functionality, and maintainability~\cite{Mantyla2009WhatReviews, Beller2014ModernFix}.
The authors can modify the submission to address the reviewers' feedback.
These activities continue until the code changes are accepted and merged into the codebase.
Nevertheless, despite the widespread code review practices, identifying security issues can be a challenge for reviewers.
Recent studies on secure code review~\cite{Braz2022SoftwarePerspective, Braz2021Why-} highlighted that manually identifying security issues in the code is difficult because reviewers may not have the required knowledge and sufficient awareness of possible issues.

\inlineheading{Vulnerability contributing commit (VCC)} is a set of code changes that encompass the modified code contributed to a vulnerability. 
It represents the subject of the code review process which aims to uncover the vulnerabilities during the development cycle.
\citet{Iannone2022TheStudy} conducted a large-scale study on the vulnerability lifecycle, revealing that real-world vulnerability can be composed of multiple commits over time.
For instance, an \textit{SQL injection} requires an average of 18 VCCs.
It can be implied that a vulnerability can potentially be prevented if the issues in VCC are identified by the reviewers, and fixed by the developers, during code review.

\inlineheading{\SATFullC{} (\SAT{})} is a quality assurance tool that identifies certain security-related issues in source code using the pre-defined set of rules without executing the program. 
We consider \SAT{s} that employ two overarching analysis techniques~\cite{Lipp2022AnDetection}: \textcircled{1} syntactic-based and \textcircled{2} semantic-based.
% It was argued that \SAT{s} may help reviewers better identify security issues~\cite{Imtiaz2019HowUsage}.
While many studies~\cite{Goseva-Popstojanova2015OnVulnerabilities, Harman2018FromAnalysis, Ernst2004StaticDuality, Sadowski2015Tricorder:Ecosystem, Aloraini2019AnTools, Penta2009TheStudy, Lipp2022AnDetection, Li2023ComparisonJava} empirically evaluated \SAT{s} on identifying security issues, they neglect the context of code reviews and VCCs.
For example, \citet{Lipp2022AnDetection} evaluated the effectiveness of tools with a partially synthetic dataset, and a recent study by \citet{Li2023ComparisonJava} evaluated the tools on the vulnerable versions of real-world programs.
% Moreover, in this work, we also evaluate \SAT{s} on a larger dataset that covers diverse types of vulnerabilities. 
The key differences between our work and recent \SAT{} studies are shown in Table~\ref{table:previous-work-comparison}.
% A study~\cite{Vassallo2020HowContexts} reported that developers regularly use \SAT{s} in two development contexts i.e., regular development and code review.
% The capability of \SAT{s} to identify security issues in the early code changes is yet to be investigated.
While previous studies related to code reviews \cite{Singh2017EvaluatingEffort, Mehrpour2022CanDefects, Panichella2015WouldReviews} investigated the capabilities of \SAT{s} in helping reviewers identify coding styles or patterns,
% the previous works that studied the \SAT{s} share some common limitations: 
there are still knowledge gaps about the capabilities of \SAT{s} for vulnerability identification.
Specifically, the following aspects have not been incorporated into the existing studies:
\textcircled{1} the context of code changes e.g., modified files and functions,
% that contain security issues that may not yet be fully exploitable but still contribute to vulnerabilities at a later stage, 
\textcircled{2} the benefits of code reviews supported by \SAT{} warnings, and 
\textcircled{3} the amount of time that reviewers need to wait for \SAT{} results.

\begin{table}[t]
    \centering
    \footnotesize
    \caption{Key differences between this work and the recent \SAT{} studies focusing on vulnerability identification. }
    \begin{tabularx}
    {\linewidth}{
            >{\hsize=1.6\hsize}X
            % >{\hsize=0.5\hsize}X
            >{\hsize=0.5\hsize}X
            >{\hsize=0.5\hsize}X
            >{\hsize=0.5\hsize}X
            >{\hsize=0.5\hsize}X
            >{\hsize=0.5\hsize}X
            >{\hsize=1\hsize}X
          }
          \toprule
% & \multicolumn{6}{c}{ \textbf{Metrics}  }\\
 \textbf{Study}
 % &  \textbf{\SAT{s}}
&  \textbf{Lang.}
 &  \textbf{Subject Projects}
&  \textbf{CVEs}
&  \textbf{CWE Groups}
&  \textbf{Exec. Time}
&  \textbf{Code Commits}
\\
\midrule
\textbf{\citet{Li2023ComparisonJava}} & 
% 7 & 
Java & 63 & 165 & 7 & \cmark &  -
% Studied seven open-source \SAT{s} using 165 Java programs, covering seven types of CWEs and report the effectiveness and execution time. & 
 \\
\textbf{\citet{Lipp2022AnDetection}} & 
% 6 &
C/C++ & 9 & 192 & 5  & \xmark & -
% Studied six free and commercial \SAT{s} using 192 C and C++ synthetic and real-world programs that contain injected vulnerabilities. \\
\\
\textbf{Our Study} & 
% 5 & 
C/C++ & 92 & 319 & 8 & \cmark & 815 VCCs
% \multirow{3}{*}{\parbox{\linewidth}{
% Study five free \SAT{s} with 815 VCCs, from 319 C and C++ CVE vulnerabilities, that cover eight types of CWEs. 
% Consider how \SAT{} can support secure code reviews by warning-based prioritization performance as well as\SAT{}'s computing time.
% }}
\\
\bottomrule
 % \multicolumn{10}{l}{\textcolor{red}{TODO - define ranges of significancy levels e.g., * 0.05, ** 0.01, *** 0.005}} \\
    \end{tabularx}
    \label{table:previous-work-comparison}
\vspace{-2mm}
\end{table}

\section{Study Design}
\label{study_design}
To bridge the knowledge gap of the \SAT{s}-supported secure code reviews, we study \SAT{s} using the \textit{real-world vulnerability contributing commits}.
% (VCC; a set of code changes that contributed to exploitable vulnerabilities).
Evaluating \SAT{s} with this dataset will help researchers and practitioners better understand the capabilities of the tools to assist reviewers in identifying potential vulnerabilities during the code review process.
We also investigate 
% code review performance when the warnings of \SAT{s} are available, as well as the 
\SAT{s}' computing time.

\begin{figure}[t]
\center
  \includegraphics[width=0.95\linewidth]{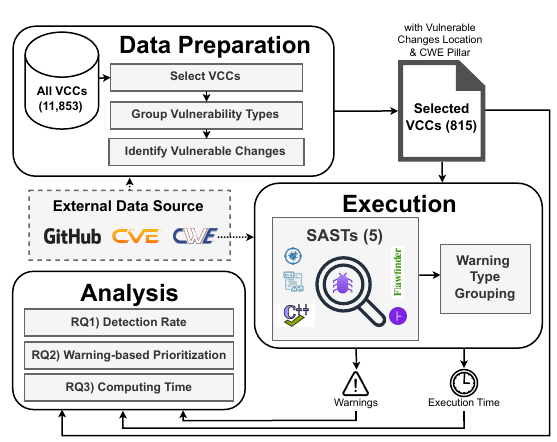}
\caption{Overview of our study approach.}
\label{fig:study-process-overview}
\vspace{-3mm}
\end{figure}

In the following, we present our research questions and the study design. The overview of the study process is shown in Figure~\ref{fig:study-process-overview}.

\subsection{Research Questions}

In this study, we set out to answer the following research questions.

\smallheading{RQ1 \RQOneTopic: \RQOne}
\underline{Motivation}: Previous work~\cite{Goseva-Popstojanova2015OnVulnerabilities, Harman2018FromAnalysis, Ernst2004StaticDuality, Sadowski2015Tricorder:Ecosystem, Aloraini2019AnTools, Penta2009TheStudy} studied \SAT{s} with synthetic benchmark datasets or certain versions of programs that contain complete vulnerabilities.
The existing datasets may not represent the vulnerabilities that gradually appear from code changes during the development process.
Investigating \SAT{s} using VCCs should extend the understanding of the tool's capability to detect vulnerabilities in this context.
% Early vulnerability detection is an objective of the security shift-left concept~\cite{Weir2022ExploringResponsibility}.
% Although many \SAT{} evaluations [-REF-] focus on the effectiveness of tools on the vulnerable version of programs, little has investigated whether \SAT{s} can detect the vulnerabilities in the contributing commits. 

\noindent\underline{Approach}: 
% To understand the vulnerability detection capability in VCCs, 
\reviewersMinorModification{We execute \SAT{s} on the selected VCCs and determine the number of VCCs that \SAT{s} can produce warnings on vulnerable code changes.}
In this work, we consider the scope of code changes in VCCs at the file and function levels.
We also analyze the types of vulnerabilities in VCCs that \SAT{s} can detect.
% We consider the number of commits that each tool can produce warnings in the following scenarios \textcircled{1} the scope of vulnerable changes (i.e., files or functions), \textcircled{2} the number of vulnerable changes that receive warnings (i.e., all changes or at least one change), and \textcircled{3} the type of warnings and the type of vulnerabilities (i.e., types match or types not match).

\smallheading{RQ2 \RQTwoTopic: \RQTwo}
\underline{Motivation}: 
Reviewers have limited review capacity~\cite{Braz2022SoftwarePerspective, Singh2017EvaluatingEffort}.
Previous work~\cite{Balachandran2013ReducingRecommendation, Beller2016AnalyzingSoftware, Heckman2008OnTechniques} argued that \SAT{} warnings may help reduce reviewer effort.
% Still, the excessive number of warnings requires more effort to inspect~\cite{Johnson2013WhyBugs, Singh2017EvaluatingEffort}.
% However, little is known about how much review effort a \SAT{} can potentially save.
\citet{Muske2016SurveyAlarms} reported that developers 
% use warning information from the tools to prioritize the code changes that they should pay attention to.
prioritize the \SAT{} warnings that they should pay attention to.
Thus, we hypothesize that warning-based prioritization of changed functions could reduce reviewing effort.
% the performance of warning-based prioritization has never been reported.

% understand the potential benefits of code review with the support of \SAT{s}.

\noindent\underline{Approach}: 
% To answer RQ2, 
% We measure warning-based prioritization performance for identifying vulnerable functions in VCCs when changed functions are ranked with \SAT{} warnings.
% \citet{Muske2016SurveyAlarms} reported that developers use warning information from the tools to prioritize the code changes that they should pay attention to.
% \citet{Muske2016SurveyAlarms} reported that developers prioritize the \SAT{} warnings to inspect, indicating that developers may also prioritize code changes that received warnings.
We set out to investigate whether, within a limited review effort, warning-based prioritization can help reviewers identify more vulnerable functions.
% We explore three prioritization~\cite{Muske2016SurveyAlarms} approaches, i.e., warning amount, warning density~\cite{Trautsch2023AreProjects}, and warning severity~\cite{Vassallo2020HowContexts}.
We measure the accuracy of identifying vulnerable functions prioritized by \SAT{} warnings within a fixed review effort (i.e., $K$\% lines of code in changed functions). 
% Inspired by a prior study that investigated the defect prediction approaches for code reviews~\cite{Hong2022WhereTo}, we calculate the performance metrics based on an assumption that reviewers rely on the warning information~\cite{Muske2016SurveyAlarms} from the tools and prioritize the changes that should be reviewed accordingly.
% We use three metrics~\cite{Wattanakriengkrai2022PredictingTechnique}, namely Precision@K, Recall@K, and Initial False Alarm to answer this question. 

\smallheading{RQ3 \RQThreeTopic: \RQThree}
\underline{Motivation}: 
% In code reviews, developers typically expect to receive feedback in a timely manner~\cite{MacLeod2018CodePractices}. 
\citet{Kudrjavets2022MiningAnalysis} suggested that automated tests should be executed during the \textit{non-productive} period of code reviews.
To incorporate \SAT{s} without delaying code reviews, the computation time should fit this idle period.
However, the computation time of C/C++ \SAT{s} on VCCs remains underexplored.
% \citet{Liu2023AJava} found that the computation time of Java \SAT{s} is related to the size of the analyzed system. 
% % Therefore, a project can estimate the time for Java \SAT{s}.
% Little is known whether this applies to other \SAT{s}.

\noindent\underline{Approach}: We measure the computation time from when the tool starts analyzing a commit until the warnings are produced. 
% We compare the execution time of each tool on the same commit to understand whether the tools require a significantly different time. 
Additionally, we analyze the computation time by the system's lines of code to understand whether the time depends on system size~\cite{Li2023ComparisonJava}.

% MOVE TO DISCUSSION POINT
% \smallheading{RQ4 \RQFourTopic: \RQFour}
% \underline{Motivation}: \SATC{s} should be combined because the single tool cannot detect all types of issues~\cite{Nunes2017OnStudy}.
% To increase the effectiveness of the \SAT{s} without sacrificing the code review effort, we aim to identify the group of \SAT{s} that offers the highest effectiveness with the lowest required manual inspection effort using our selected VCCs.

% \noindent\underline{Approach}: [TBD]

\subsection{Data preparation}
\label{dataset_preparation}
% To develop the novel dataset for this study, we investigate code changes that potentially introduced vulnerabilities. 
Since the benchmark datasets from prior studies~\cite{Lipp2022AnDetection, Goseva-Popstojanova2015OnVulnerabilities} 
% are synthetic (i.e., injected vulnerabilities) or 
are obtained from released versions, they may not represent the nature of code changes in the software development and code review processes where the changes are small and may be mixed with other changes.
Thus, a new dataset of code changes that contributed to exploitable vulnerabilities is more suitable. 
We describe the data preparation process for our study below.

\subsubsection{Dataset} \label{explain-vccs}
% Iannone et al. \cite{Iannone2022TheStudy} conducted a large-scale study to understand the existence of vulnerabilities in a broad range of projects. They collected publicly reported vulnerabilities and applied the SZZ algorithm~\cite{SliwerskiJacek2005WhenFixes} to identify the earliest commits that introduced each vulnerability.
To craft the dataset, we begin with the Vulnerability Contributing Commits (VCCs) dataset provided by \citet{Iannone2022TheStudy}.
This dataset is suitable for our study because it collected real-world code commits that have been identified as contributing to exploitable vulnerabilities.
The VCC datasets were collected from GitHub projects that are associated with vulnerabilities in the Common Vulnerabilities and Exposures (CVE) database~\cite{CVEOverview}---the collection of exploitable vulnerabilities in software systems that are publicly reported by the community.
Given a fixing commit recorded in a CVE, a VCC was identified using the SZZ algorithm~\cite{SliwerskiJacek2005WhenFixes}.
% The original dataset consists of Vulnerability Contributing Commits (so-called VCCs) on public GitHub projects that were obtained from the Common Vulnerabilities and Exposures (CVE) database~\cite{CVEOverview}, which is the collection of exploitable vulnerabilities in software systems that are publicly reported by the community. 
Each VCC includes meta-information, i.e., the repository URL, commit identification number (Commit SHA), related vulnerability (CVE number), and related vulnerability type (CWE item).
It should be noted that the VCC dataset does not provide the actual changes in each commit, e.g., vulnerable files or vulnerable functions.
% Below, we describe each step of our data preparation approach.
% The demographics of the original VCC dataset and the selected VCC dataset for this study are shown in Table~\ref{table:original-vcc-dataset}.
% The data preparation procedure is explained in Section~\ref{vcc_selection} and Section~\ref{identify_vulnerable_changes}.

% \input{table/table-original-vcc-dataset}

\subsubsection{Selecting VCCs}
\label{vcc_selection}
Although the VCC dataset of \citet{Iannone2022TheStudy} is sizeable, some VCCs cannot be analyzed by \SAT{} because of incompatible programming languages or compilation issues; others may have insufficient information for our analysis (e.g., missing vulnerability type).
We select VCCs based on the following criteria:

\begin{enumerate}
\item \textbf{From C/C++ projects:} We selected VCCs from the projects implemented in C or C++ because these languages allow the programs to interact with low-level operations that can lead to exploitable vulnerabilities~\cite{Turner2014SecurityRuby}.
To determine this, we obtained the project's main programming language from GitHub API.

\item \textbf{Have an assigned vulnerability type:} Since we will analyze the vulnerability type that \SAT{s} can detect, we need to ensure that the selected VCCs have sufficient vulnerability information.
The associated CVE of a VCC is usually assigned a vulnerability type in the Common Weakness Enumeration (CWE) taxonomy.
Thus, we selected VCCs that have at least one CWE item assigned to the associated CVE.

\item \textbf{Compilable:}
Some \SAT{s} require the information from the compilation process.
% of a VCC.
Thus, the selected VCCs must be compilable.
% However, it is not feasible to test the compilation of all VCCs.
\reviewersMinorModification{
However, the presence of diverse build processes and dependencies across many C/C++ projects in the VCC dataset poses challenges for automated compilation tests, increasing the chance of build failure.
Manually compiling every VCC is also infeasible due to the large number of VCCs.
To mitigate the issue,}
we first selected the projects that use the standard build process i.e., \texttt{GNU Autotools} as it has been used by many C and C++ projects (over 38\%) in the original VCC dataset.
Then, we selected VCCs that can be compiled successfully following the instructions in the project's document.

\end{enumerate}
Based on the first criterion, we obtained 5,354 VCCs in 341 projects from the VCC dataset.
Then, using the second criterion for the obtained VCCs, we identified 5,015 VCCs linked to 1,654 CVEs that have at least one CWE item.
Finally, 1,057 VCCs with 371 CVEs from 92 projects passed the compilation test in our third criterion.

% To prepare the dataset for our study, we select VCCs from original VCC dataset based on the constraints: \textcircled{1} VCCs are from the project that uses C and C++ as the main programming language, \textcircled{2} types of vulnerabilities can be mapped to CWE taxonomy, \textcircled{3} VCCs can be compiled using the procedure, and \textcircled{4} location of vulnerable changes can be identified. 
% % The selection process is illustrated in Figure \textcolor{red}{xxx}.

% The size of VCCs selected for this study are shown in Table~\ref{table:selected-vcc-size}. The types of vulnerabilities from our selected VCCs are shown in Table~\ref{table:selected-vcc-vulnerability-type}

\begin{table}[t]
\centering
\footnotesize
\caption{Vulnerability types of selected VCCs. A single VCC can be related to more than one CWE pillar.}
\begin{tabularx}{\linewidth}{
            >{\hsize=0.7\hsize}X
            >{\hsize=2.2\hsize}X
            >{\hsize=0.05\hsize}X
          }
\toprule
\textbf{Short Name} & \textbf{CWE Pillar} & \textbf{\#VCCs}\\
 \midrule
 \textbf{Access Ctrl} & \textbf{CWE-284} Improper Access Control$^\dag$ & 38 \\ 
 \textbf{Resource Ctrl} & \textbf{CWE-664} Improper Control of a Resource Through its Lifetime & 538 \\ 
 \textbf{Incorrect Cal} & \textbf{CWE-682} Incorrect Calculation & 78 \\ 
 \textbf{Control Flow} & \textbf{CWE-691} Insufficient Control Flow Management & 56 \\ 
 \textbf{Protect Mech} & \textbf{CWE-693} Protection Mechanism Failure$^\dag$ & 13 \\ 
 \textbf{Cond Check} & \textbf{CWE-703} Improper Check or Handling of Exceptional Conditions & 20 \\ 
 \textbf{Neutalization} & \textbf{CWE-707} Improper Neutralization & 56 \\ 
 \textbf{Coding Std} & \textbf{CWE-710} Improper Adherence to Coding Standards$^\dag$ & 45 \\ 
\bottomrule
\multicolumn{3}{l}{$\dag$ indicates the additional pillars from \cite{Lipp2022AnDetection}} \\
\end{tabularx}
\label{table:selected-vcc-vulnerability-type}
\end{table}
\vspace{-2mm}

\subsubsection{Vulnerability Type Grouping}
\label{sec:vul_type_grouping}
\begin{figure}[h]
\center
  \includegraphics[width=0.9\linewidth]{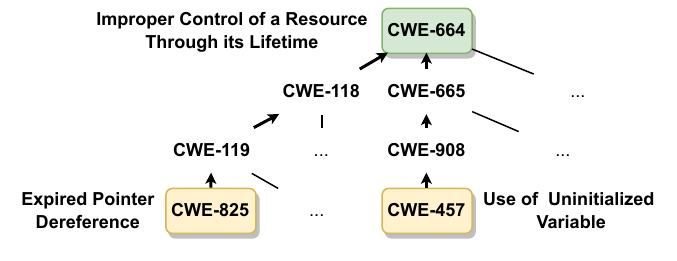}
\caption{An illustrative example of grouping CWE items (CWE-825 and CWE-457) to the CWE pillar (CWE-664).}
\label{fig:cwe-tree-example}
\end{figure}

We used vulnerability information in a \textit{CWE item} to analyze the vulnerability type in this study.
CWE items represent certain weaknesses that can lead to vulnerabilities in a software system.
However, while the CWE items enable detailed analyses of individual vulnerabilities, they do not embody the similar vulnerabilities in the VCC dataset that are assigned with different CWE items.
Therefore, we grouped the assigned CWE items of VCCs to a \textit{CWE pillar}, which is the top level of the CWE hierarchy, representing abstract categories of the CWE items~\cite{Goseva-Popstojanova2015OnVulnerabilities}.
\reviewersMinorModification{Note that we do not use other CWE grouping methods such as CWE categories~\cite{2023CWE4.12} because they only cover 400 CWE items out of over 900 CWE items in the ten pillars of CWE hierarchy.}
To identify the associated pillar of a CWE item, we traced the CWE item's parent to the root of the tree.
Figure \ref{fig:cwe-tree-example} illustrates our grouping approach.
For example, CWE items \texttt{Use of Uninitialized Variable} (CWE-457)~\cite{2023CWE4.12b} and \texttt{Expired Pointer Dereference} (CWE-825)~\cite{2023CWE4.12c} will be grouped into the \texttt{Improper Control of a Resource Through its Lifetime} (CWE-664) pillar~\cite{2023CWE4.12d}.
% because CWE-664 is the parent of CWE-118 and CWE-665, which are the grandparent of CWE-825 and CWE-457 via CWE-119 and CWE-908 respectively.
% CWE~\cite{CWEOverview} is the taxonomy that classifies the types of vulnerabilities in the hierarchical structure.
% CWE items represent certain weaknesses in the software development process which can lead to vulnerabilities in the software systems. 
% CWE pillars sit at the top level of the CWE hierarchy and represent the abstract categories of weaknesses that a CWE item belongs to~\cite{Goseva-Popstojanova2015OnVulnerabilities}.

% For each of the selected VCCs, we determined the CWE pillar based on the assigned CWE item.
In total, our dataset covers eight CWE pillars, which is more diverse than the benchmark dataset of prior work~\cite{Lipp2022AnDetection} (Table \ref{table:previous-work-comparison}).
Table \ref{table:selected-vcc-vulnerability-type} shows the number of VCCs in each of the CWE pillars.

\begin{table}[t!]
\centering
\footnotesize
\caption{Demographics of code changes in selected VCCs. 
The mean and median values compare the central tendencies between each pair of items.
% \textcolor{red}{Note that vulnerable changes are small compared to total changes.}
}
\begin{tabularx}{\linewidth}{
            >{\hsize=1.8\hsize}X
            >{\hsize=0.4\hsize}X
            >{\hsize=0.4\hsize}X
            >{\hsize=1.4\hsize}X
          }
\toprule
\textbf{Description} & \textbf{Mean} & \textbf{Median } & \textbf{Distribution}\\ 
 \midrule
 \multicolumn{4}{@{} l}{\textbf{CVE Vulnerabilities (319)}} \\
    VCCs Contributed to CVE & 2.5 & 2 & 
    \parbox[c]{1em}{
    \includegraphics[width=1in]{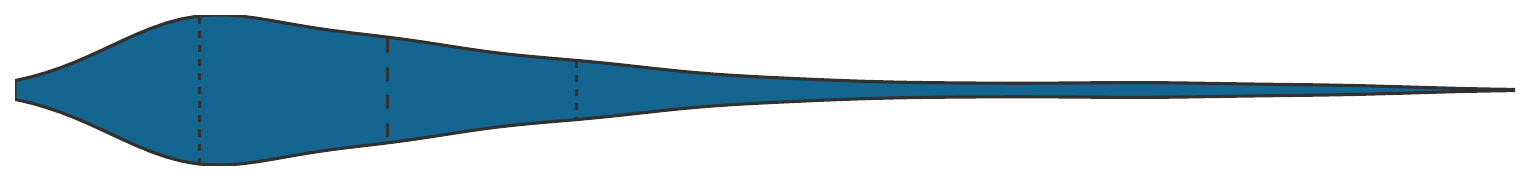}}\\
 \hline
 \multicolumn{4}{@{} l}{\textbf{VCCs with Vulnerable Files (815)}} \\
    Changed Files (9,851) & 12 & 3 &
    \multirow{2}{*}{
        \parbox[c]{1em}{
        \includegraphics[width=1in]{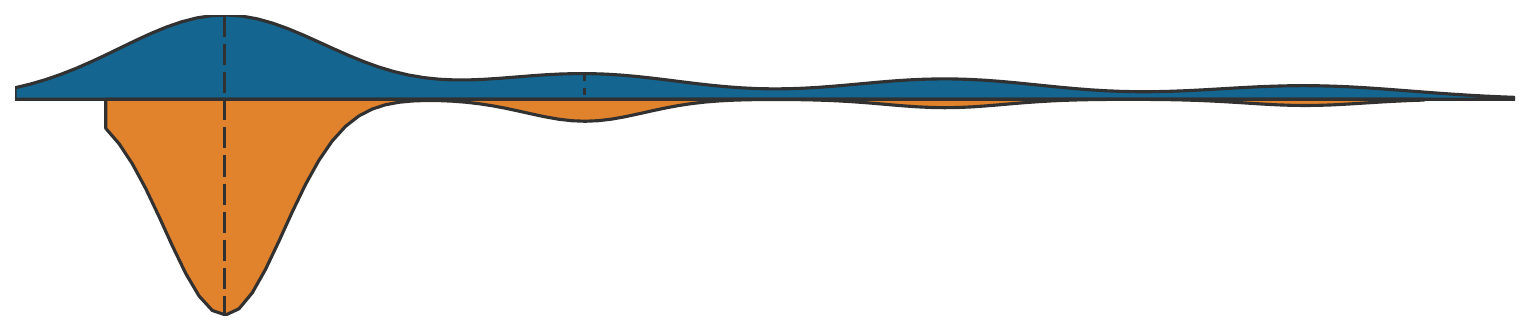}}
    }
    \\
    Vulnerable Files (1,064) & 1 & 1 & 
    % \parbox[c]{1em}{\includegraphics[width=1in]{figure/vcc_demographic/vulnerable_files.pdf}} 
    \\
    \hline
    LOC in Changed Files & 9,956 & 2,492 & 
    \multirow{2}{*}{
        \parbox[c]{1em}{
        \includegraphics[width=1in]{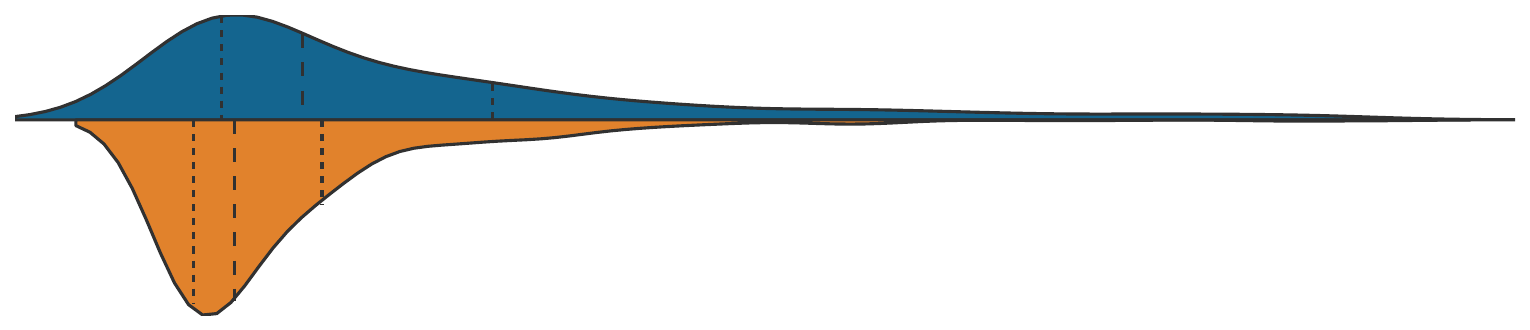}}
    }
    % \parbox[c]{1em}{\includegraphics[width=1in]{figure/vcc_demographic/loc_changed_files.pdf}} 
    \\
    LOC in Vulnerable Files & 1,820 & 1,034 & 
    % \parbox[c]{1em}{\includegraphics[width=1in]{figure/vcc_demographicloc_vulnerable_files.pdf}} 
    \\
 \hline
 \multicolumn{3}{@{} l}{\textbf{VCCs with Vulnerable Functions (697)}} \\
    Changed Functions (34,541) & 38.9 & 6 & 
     \multirow{2}{*}{
        \parbox[c]{1em}{
        \includegraphics[width=1in]{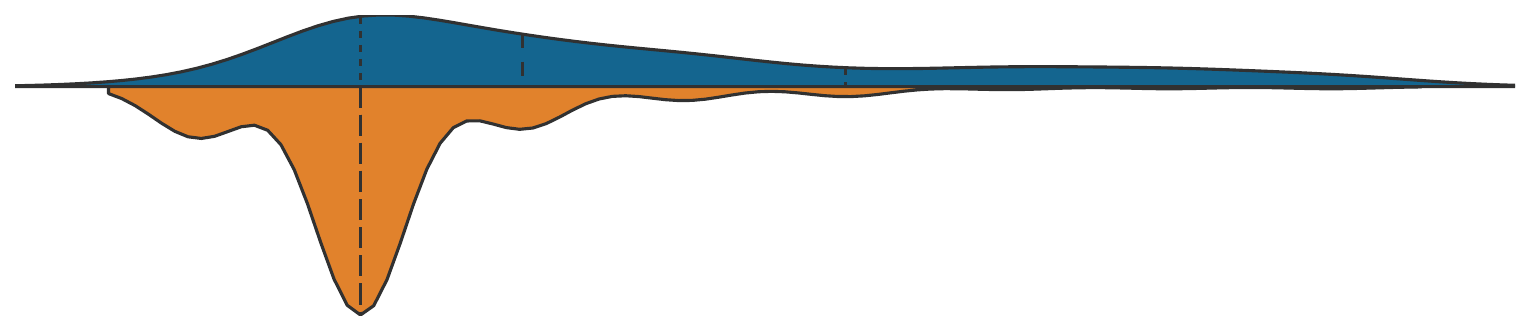}}
    }
     % \parbox[c]{1em}{\includegraphics[width=1in]{figure/vcc_demographic/changed_functions.pdf}} 
     \\
    Vulnerable Functions (1,060) & 1.5 & 1 & 
    % \parbox[c]{1em}{\includegraphics[width=1in]{figure/vcc_demographic/vulnerable_functions.pdf}} 
    \\
    \hline
    LOC in Changed Functions & 3,304 & 531 & 
     \multirow{2}{*}{
        \parbox[c]{1em}{
        \includegraphics[width=1in]{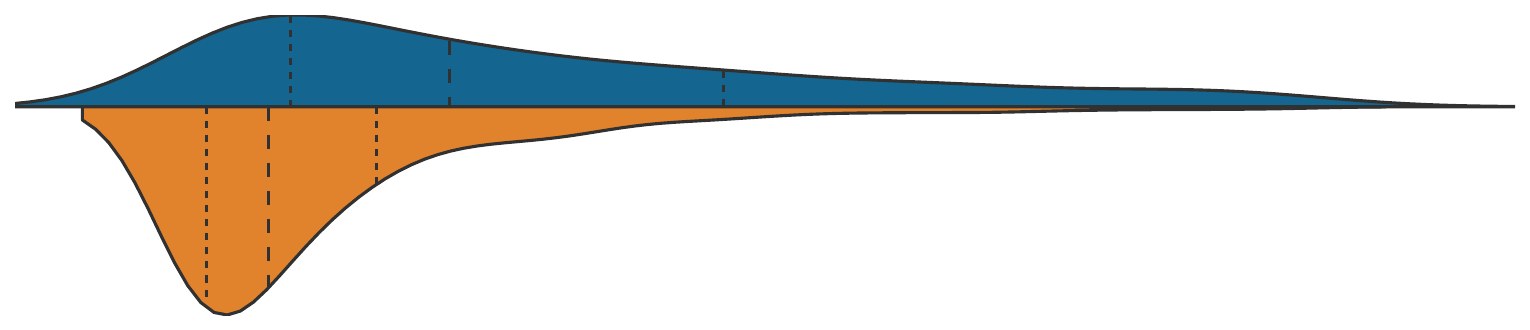}}
    }
     % \parbox[c]{1em}{\includegraphics[width=1in]{figure/vcc_demographic/loc_changed_functions.pdf}} 
     \\
    LOC in Vulnerable Functions& 221 & 114 & 
    % \parbox[c]{1em}{\includegraphics[width=1in]{figure/vcc_demographic/loc_vulnerable_functions.pdf}} 
    \\

 \bottomrule
\end{tabularx}
\newline
\label{table:selected-vcc-dataset}
\end{table}
\vspace{-2mm}

\subsubsection{Identifying vulnerable changes location in VCCs}
\label{identify_vulnerable_changes}
% Not all code changes in a VCC contribute to vulnerabilities. 
Since we aim to investigate whether \SAT{s} can provide warnings to the right locations of vulnerable code, we must identify the scope of vulnerable code in a VCC.
We considered two granularity levels, i.e., changed files and changed functions, in VCCs.
% Due to SZZ algorithm's limitations~\cite{Rosa2021EvaluatingOracle}, 
\reviewersMinorModification{
We did \textit{not} identify the vulnerable changes at the line level because modified lines in fixing commits may not directly correspond to vulnerable lines in VCCs~\cite{Rosa2021EvaluatingOracle}, hence the reliability is reduced.
However, we also manually evaluated \SAT{} warnings to understand their actual relevancy to the VCCs in subsequent analysis (see Section~\ref{tool_performance_discussion}).}

Using similar intuition as in previous work~\cite{SliwerskiJacek2005WhenFixes}, we considered that a changed file (or a changed function) is vulnerable if it was modified in the associated vulnerability-fixing commits.
% To identify the vulnerable changes in VCC, we automatedly annotate the changes by examining the commits that fixed the vulnerability.
Following the approach of~\cite{Iannone2022TheStudy}, we obtained the fixing commits from GitHub hyperlinks in the CVE records of VCCs.
% We collect GitHub links in the vulnerability's CVE record which designate the commits that fix the vulnerability.
% Then, we match the changes in VCCs with the changes in fixing commits in two granularity levels, namely file-level and function-level.
The changed files in VCCs are considered vulnerable (\textit{\textbf{vulnerable files}}, henceforth) when these files were also changed in the fixing commits. 
For the changed functions, we used \texttt{lizard} package~\cite{2024Terryyin/lizard:Languages.} to identify changed functions in a VCC and the associated fixing commits.
The changed functions in VCCs are considered vulnerable (\textit{\textbf{vulnerable functions}}, henceforth) when these functions were also changed in the fixing commits.
% A VCC is considered vulnerable at a level (i.e., file and function) if the changes in that level are later modified in the fixing commits i.e., the same file is modified in VCCs and in fixing commit or the same function is modified in VCCs and in fixing commits.
% In particular, we use \texttt{lizard} package\footnote{\url{https://github.com/terryyin/lizard}} in Python to identify the functions.
% Using this procedure, we discard \textcolor{red}{xxx} VCCs.

We successfully located vulnerable code changes in
815 VCCs where at least one source code file is modified in both VCCs and fixing commits. 
These VCCs contributed to 319 exploitable vulnerabilities in the CVE database.
Table~\ref{table:selected-vcc-dataset} shows the demographics of selected VCCs, emphasizing that 
% \textcircled{1} a vulnerability usually involves multiple VCCs and \textcircled{2} 
the vulnerable changes can be difficult to identify because they are typically small compared to all code changes in each VCC.
We use these VCCs in our analysis.

% [..] observation - Realistic vulnerabilities are not atomic. They can be introduced by multiple code changes, and fixed by multiple commits.

\label{analysis}
\subsection{Execution}
This section describes the studied \SAT{s}, experiment setup, and warning type grouping.

% methodology by explaining detection scenarios, metrics, and infrastructure.

% \subsection{Overview}

% Why combining SAT - Fuzzing: duality / synergy idea \cite{Tudela2020OnApplications, Ernst2004StaticDuality}

% To analyze the \SAT{s}, we execute tools on the selected VCCs and collect the results (i.e., warnings and computation time from each commit) for further analysis. 
% Figure \textcolor{red}{xxx} shows the overview of our analysis process.

% \begin{figure}[h]
% \center
%   \includegraphics[width=\linewidth]{figure/overall-evaluation-process.pdf}
% \caption{Overall analysis process}
% \label{fig:analysis-process}
% \end{figure}

\subsubsection{Studied Tools} \label{tool_selection_criteria}
We aim to analyze \SAT{s} that are generally available for C and C++ OSS projects. 
% slightly similar to [\textcolor{red}{Li et al. - FSE}] 
Specifically, we select \SAT{s} that are 
\textcircled{1} available for use free of charge,
% regardless of whether the tool is proprietary or open-source, 
\textcircled{2} able to run on the command-line interface (CLI) as we need to automate the compilation process, 
and \textcircled{3} actively maintained by the developers (been updated in the \reviewersMinorModification{last 12 months}).
% and \textcircled{4} provide documentation for the vulnerability type analysis.

To select \SAT{s}, we first obtain 53~\SAT{s} from prior studies~\cite{Stefanovic2020StaticReview, Lipp2022AnDetection, Nachtigall2022ATools} and NIST's Software Assurance Metrics And Tool Evaluation (SAMATE)~\cite{2023SourceNIST} as candidates.\footnote{The exhaustive list of assessed \SAT{s} is included in the data package}
Then, we examine the documentation of each tool to understand its specifications.
From the initial list, we remove 12~\SAT{s} that do not support C and C++ languages.
Then, we exclude 26 commercial~\SAT{s}, two~\SAT{s} that do not operate on the command-line interface, and four~\SAT{s} that are not actively maintained.
\reviewersMinorModification{
We also omit four \SAT{s}, including those that operate on an external server, are compiler extensions (not employing further rules), or lack information on available warnings.
}
Finally,
three semantic-based tools, i.e., 
CodeChecker\cite{2024CodeChecker}, 
CodeQL~\cite{2024CodeQL}, and
Infer~\cite{2023InferAnalyzer};
one syntactic-based tool, i.e.,
Flawfinder~\cite{2023Flawfinder};
and one hybrid tool, i.e.,
Cppcheck~\cite{2024CppcheckAnalysis}
pass our criteria.
% The versions of \SAT{s} used in this study are shown in Table~\ref{table:tool-version}. 
% The exhausted list of reviewed tools is available in the supplementary package.

% \input{table/table-tool-version}

\subsubsection{Experiment Setup} 
% \inlineheading{Setup} - 
In this study, we use the latest stable versions of the selected \SAT{s}, i.e., Cppcheck v2.10, CodeChecker v6.22.2, CodeQL v2.13.3, Flawfinder v2.0.19, and Infer v1.10.
% We review the configurable parameters of each tool to ensure that the optimal settings are used in our experiments. 
For each tool, we use the settings recommended by its developers for reliable results (as also suggested by~\cite{Metzman2021FuzzBench:Service}).
In particular, for Flawfinder and Cppcheck, we use the readily enabled rules.
For CodeQL, we use the LGTM query suite~\cite{2021StandardC/C++}, which contains the largest number of rules.
For CodeChecker and Infer, we use the recommended settings. 
Note that we do not enable all rules for CodeChecker and Infer as it is discouraged by the tools' developers.\footnote{\url{https://codechecker.readthedocs.io/en/latest/analyzer/user_guide/\#enable-all} and \url{https://github.com/facebook/infer/issues/1114\#issuecomment-506284374}}

\inlineheading{Pipeline} - We develop an automated framework to facilitate \SAT{} execution. 
Each VCC is downloaded and prepared before running \SAT{}. 
% The execution results (i.e., execution status, computation time, and number of warnings) are collected in a centralized database. 
The preparation steps are different between \SAT{s}.
For \SAT{s} that require compilation (i.e., CodeChecker, CodeQL, and Infer), we update \texttt{Makefile} by running the automated script that the projects have prepared for the compilation, i.e., \texttt{autogen.sh}, \texttt{bootstrap.sh}, and \texttt{build.sh}. If none of the scripts is present, we use the \texttt{autoreconf} command to generate the required files.
Then, we run the configuration file, i.e., \texttt{configure}.
% \reviewersExclusion{
% On the other hand, Flawfinder requires all source files to be encoded in UTF-8. Thus, we used the \texttt{ftfy} Python package~\cite{Speer2019Ftfy} to convert source code files to UTF-8.}
After the required preparation, we execute \SAT{} and collect the produced warnings for further analysis.
Note that the time spent on preparation is also included in the computation time.

% To support future works that conduct large-scale \SAT{} studies on the open-source code commits, we release this pipeline with detailed instructions~\cite{Anonymous2023AutomatedCommits}.
The pipeline operates in isolated Docker containers to control the environment and resources for each experiment (i.e., automatically running \SAT{} on a selected set of target code commits).
It stores the produced warnings on the host machine for ease of access and manages a centralized database that collects execution status, number of warnings, and timestamps for further analysis.
% Additional \SAT{s} can be integrated with exclusively customized configurations.
% \reviewersExclusion{Additional \SAT{s} can be integrated with exclusively customized configurations.}

% Finally, we compile the program using \texttt{make -i} command. Note that we also use this procedure for \SAT{} execution.}
% Analysis artifacts (i.e., reports or warning files) are stored separately for further analysis.
% We find that Flawfinder cannot analyze a large portion of selected VCCs due to the incompatible encoding in source code files.

\inlineheading{Infrastructure} - 
To run our experiments, we use a virtual machine with a 32-core virtual CPU and 288 GB of memory. 
We run each of the studied \SAT{s} in an Ubuntu 22.04 Docker container with a 4-core virtual CPU.
% To prevent the unrealistic computation time, 
Due to the time constraint, we terminate an execution when it continues for more than five hours on any VCCs.
% , which is a reasonably unrealistic computation time.

\subsubsection{Warning Type Grouping} 
\label{warning_type_mapping}
Since we aim to examine whether a \SAT{} provides warnings that match the vulnerability type of a VCC, we mapped and grouped warnings of the studied \SAT{s} into CWE pillars.
For Cppcheck, CodeQL, and Flawfinder, each of their warnings has a CWE item assigned.
Thus, we used the same approach as described in the vulnerability grouping to group the CWE items into CWE pillars (see Section \ref{sec:vul_type_grouping}).

Two \SAT{s}, i.e., CodeChecker and Infer, do not assign CWE items in the warnings.
Thus, the warnings produced by these tools must be mapped to the most relevant CWE pillars.
We used the mapping between warnings and CWE items provided by~\citet{Lipp2022AnDetection}.
However, the mapping~\cite{Lipp2022AnDetection} does not cover all warning types in our VCCs which contain more diverse vulnerabilities.
We consult the official documents~\cite{2024ClangCheckers, 2022Clang-TidyCheckers, 2024InferTypes} to map the remaining warnings without CWE items.
For example, we assign CWE item \textit{Reachable Assertion} (CWE-617) to the \texttt{bugprone-assert-side-effect}\footnote{\url{https://clang.llvm.org/extra/clang-tidy/checks/bugprone/assert-side-effect.html}} warning of CodeChecker.
% as it checks for the reachable \texttt{assert()} function.
Out of 178 warnings that needed the new mapping, 5\% cannot be linked to any CWE items.\footnote{We include the complete mappings in our data package}
% The mapping process was conducted by the first author and reviewed by the third author who has over 10 years of experience in software security testing.
\reviewersMinorModification{The mapping process is conducted primarily by the first author.
To ensure the accuracy of the mapping process, the third author, who has over 10 years of experience in software security testing, joined the first author to establish the foundational framework and assisted with ambiguous warnings encountered along the process}.
Finally, the CWE items are grouped into the CWE pillar as described in Section~\ref{sec:vul_type_grouping}.

\section{Analyses \& Results} 
\label{results}
We report the results of our analysis and answer the research questions in this section.

\subsection{Detection~\RQOneTopic{} (RQ1)} 
\smallheading{Approach}
To answer our RQ1: \textit{\RQOne}, we measure how many VCCs that \SAT{s} can \reviewersModification{produce the warnings on the vulnerable changes} as well as analyze the types of vulnerabilities.
Specifically, we execute the studied \SAT{s} on the 815 selected VCCs and examine whether the tools produce warnings to the vulnerable file and vulnerable functions in the VCCs.
Similar to prior works~\cite{Lipp2022AnDetection, Li2023ComparisonJava}, 
for each granularity (i.e., file or function level), we consider the detection \reviewersModification{rate} based on four scenarios (see Table~\ref{table:describe-detection-scenarios}).

\begin{table}[t]
\centering
\footnotesize
\caption{Name and description of detection scenarios in two change levels.}
\begin{tabularx}{\linewidth}{
            >{\hsize=0.6\hsize}X
            >{\hsize=0.6\hsize}X
            >{\hsize=1.8\hsize}X
          }
\toprule
  % \multirow{2}{*}{\textbf{Category (CWE-\#)}} & \multicolumn{2}{c|}{\textbf{OpenSSL}} & \multicolumn{2}{c}{\textbf{PHP}} \\
  %       & \# & \% & \# & \% \\

  \multicolumn{2}{c}{\textbf{Scenario Name by Change Level}} & \textbf{Description} \\
  \cline{1-2}
  \textbf{File} & \textbf{Function} & \\
  \hline

    S1:\textit{1F-Any} &  S5:\textit{1Fn-Any} & \underline{At least one} vulnerable \textit{change} in the VCC receives warnings with \underline{any} type. \\     
    \hline
    S2:\textit{1F-Same} &  S6:\textit{1Fn-Same} & \underline{At least one} vulnerable \textit{change} in the VCC receives warnings with the \underline{same} vulnerability type as the VCC. \\     
    \hline
    S3:\textit{allF-Any} &  S7:\textit{allFn-Any} & \underline{All} vulnerable \textit{changes} in the VCC receive warnings with \underline{any} type. \\     
    \hline
    S4:\textit{allF-Same} &  S8:\textit{allFn-Same} & \underline{All} vulnerable \textit{changes} in the VCC receive warnings with the \underline{same} vulnerability type as the VCC. \\     
    
 \bottomrule
\end{tabularx}
\newline
\label{table:describe-detection-scenarios}
\end{table}

% \vspace{-5mm}

% In this work, we considered the scope of vulnerable code in VCCs at the changed files and changed functions.
% We also analyzed the types of vulnerabilities in VCCs.

\begin{figure}[t!]
\center
  \includegraphics[width=\linewidth]{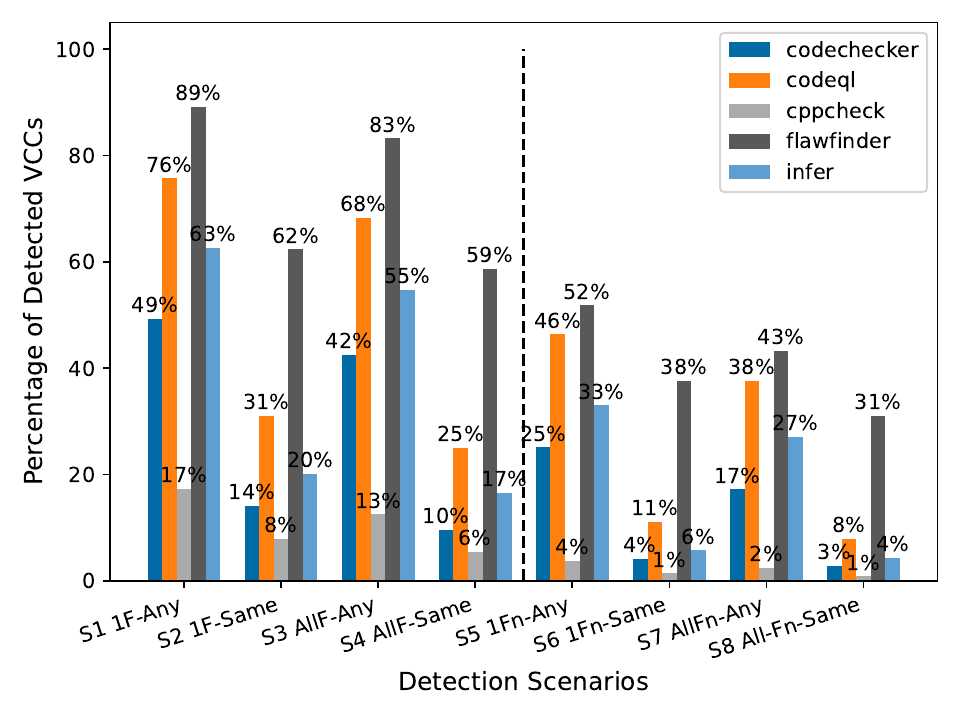}
\caption{Percentages of VCCs that the tools can detect in different scenarios.}
\label{fig:vcc-detection-scenarios}
\end{figure}

\label{rq1_result}
\smallheading{Results}
\inlineheadingNoSpace{Detection rate:} Figure~\ref{fig:vcc-detection-scenarios} shows the percentages of VCCs that an \SAT{} can detect in each scenario.
% Flawfinder consistently achieves the highest detection rate in all eight scenarios, followed by CodeQL and Infer.
\reviewersModification{Flawfinder produced warnings in vulnerable files and functions of the largest number of VCCs. Hence, it potentially offers the lower false negatives i.e., a low number of vulnerable files and functions that do not receive a warning.}
Specifically, Flawfinder produces warnings in at least one vulnerable file (S1:\textit{1F-Any}) for 89\% of the VCCs.
At the function level, which has a smaller scope of code changes, Flawfinder produces warnings in at least one vulnerable function (S5:\textit{1Fn-Any}) in 52\% of VCCs.
\reviewersModification{In contrast, Cppcheck produced warnings in vulnerable files and functions of the smallest number of VCCs in all eight scenarios.}

\textbox{\textbf{Finding (\reviewersModification{detection rate})}: Flawfinder can produce warnings in the vulnerable changes in 89\% of the VCCs at the file level and in 52\% of the VCCs at the function level.}

\begin{figure}[t!]
\center
  \includegraphics[width=0.7\linewidth,trim = 0 180 0 150, clip=true]{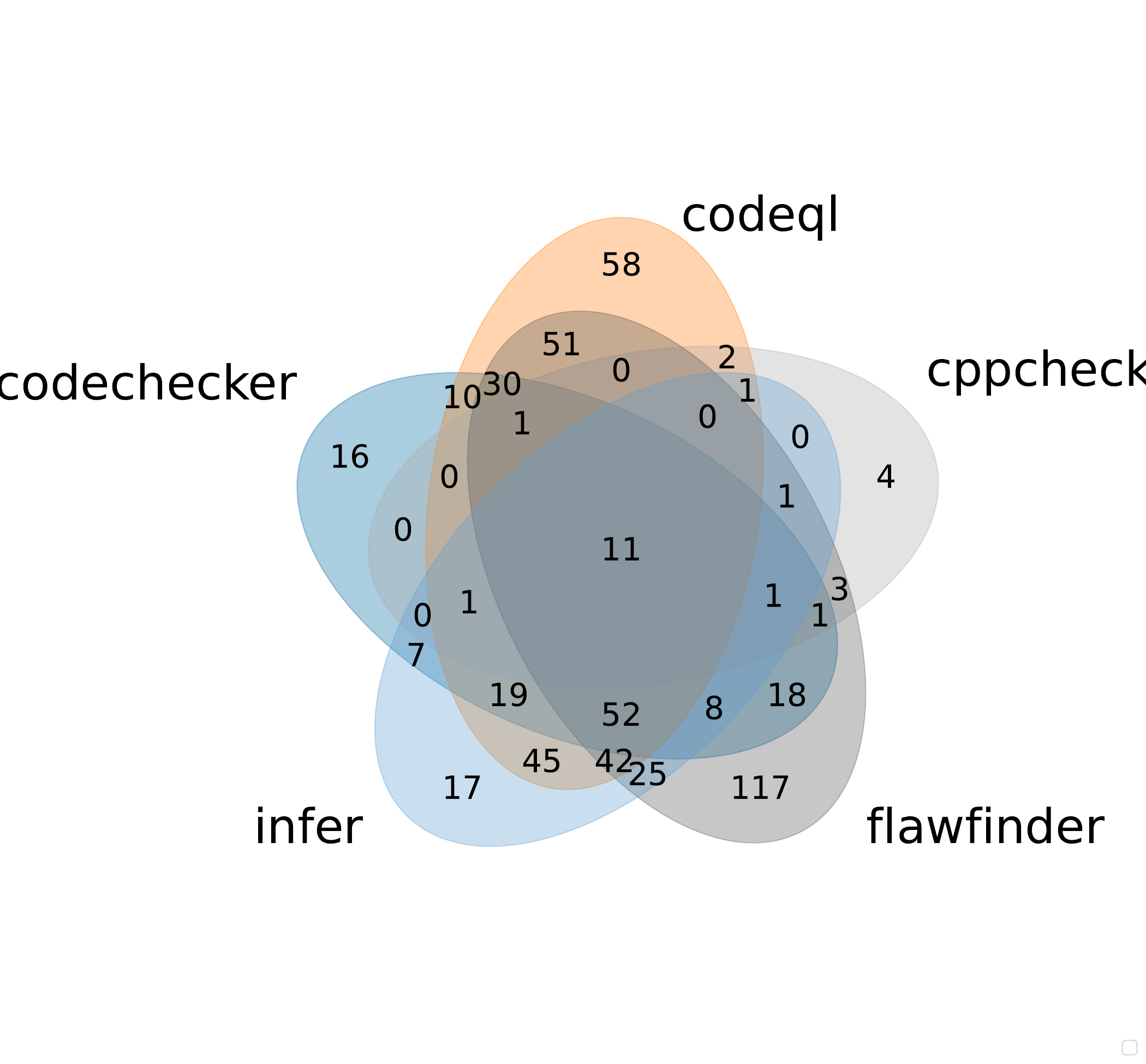}
\caption{A Venn diagram displaying VCCs for which the tools can produce warning(s) in the vulnerable functions (S5:\textit{1Fn-Any})}
\label{fig:tool-vcc-venn}
\end{figure}

We further investigate whether combining \SAT{s} can improve the detection rate.
% the effectiveness of vulnerability identification in VCCs.
% \textcolor{red}{as it was recommended that several tools should be used together to harness the cumulative benefits~\cite{Beller2016AnalyzingSoftware}}.
% \citet{Nunes2017OnStudy} also found that single \SAT{} can be more effective in identifying vulnerabilities in some scenarios~.
It is possible that different \SAT{s} can detect different vulnerability types.
% To investigate such an argument, 
We examine the detection rate of multiple \SAT{s} at the function level  \textit{(S5: 1Fn-Any)} as~\citet{Lipp2022AnDetection} remarked that function-level is an appropriate granularity for analyzing \SAT{} warnings.
% Additionally, relaxing the warning types also reduces the chance of overconstraint. 
% The number of unique VCCs that the tools can detect is shown in Figure~\ref{fig:tool-vcc-venn}.
% Figure~\ref{fig:tool-vcc-venn} shows the unique VCCs that tools can detect in scenario S5 (\textit{1Fn-Any}).
Figure~\ref{fig:tool-vcc-venn} shows that multiple tools (i.e., CodeChecker, CodeQL, Cppcheck, and Flawfinder) can commonly detect 52 VCCs, which account for only 6\%. 
% Despite being the largest group, it still represents only 6\% of the entire dataset,
This suggests that the tools do not regularly detect the same VCCs.
In that respect, when all warnings of the five tools are considered, they collectively detect 541 VCCs (78\%) which is substantially increased by 26 percentage points compared to the detection rate in \reviewersModification{S5:\textit{1Fn-Any} of a single tool with the highest detection rate (i.e., Flawfinder)}.
This highlights that combining these tools enhances their effectiveness in identifying a larger number of VCCs.

% \vspace{-10mm}

\textbox{\textbf{Finding (combining tools)}: Combining multiple \SAT{s} can increase the effectiveness in identifying vulnerability in VCCs by 26\%.}

% Vulnerability type perspective
\inlineheading{Detected vulnerabilities by type:} \SAT{s} can produce warnings on VCCs of almost all vulnerability types in our dataset. 
Table~\ref{table:detected-vcc-vulnerability-type} shows the numbers and percentages of VCCs that \SAT{s} produce warnings with the \reviewersModification{matching} type of vulnerability in at least one vulnerable function (S6:\textit{1Fn-Same}) and all vulnerable functions (\textit{S8: allFn-Same}) by CWE pillars. 
% Except for vulnerability type \textbf{Resource Ctrl} (CWE-664), every tool can only warn a small number of VCCs related to all types of vulnerabilities.
% \reviewersModification{Table~\ref{table:detected-vcc-vulnerability-type} shows that each \SAT{} can produce different warning types.}
% The examples of vulnerabilities that each tool can produce warnings on most VCCs are as follows:
\begin{itemize}
    \item \textbf{Flawfinder} warns most VCCs related to \textbf{Resource Ctrl} (CWE-664) (38\%), e.g., possible buffer overflow~\cite{2018NVDCVE-2018-7186}
    % \footnote{
    % % \url{https://github.com/danbloomberg/leptonica/commit/a9adf23592525a7cf77d23a62ffafc7ae5677aaf}
    % \url{https://nvd.nist.gov/vuln/detail/CVE-2018-7186}
    % }
    and \textbf{Neutralization} (CWE-707) (16\%), e.g., potential command injection from the insufficient input validation~\cite{2015NVDCVE-2015-9059}.
    % \footnote{
    % % \url{https://github.com/npat-efault/picocom/commit/464e25056f19ce1c486e35ad969d8b5928915d4d}
    % \url{https://nvd.nist.gov/vuln/detail/CVE-2015-9059}
    % } 
    \item \textbf{CodeQL} warns most VCCs related to \textbf{Access Ctrl} (CWE-284) (3\%), e.g., potential information leak~\cite{2020NVDCVE-2020-5260}
    % \footnote{
    % % \url{https://github.com/git/git/commit/abca927dbef2c310056b8a1a8be5561212b3243a}
    % \url{https://nvd.nist.gov/vuln/detail/CVE-2020-5260}
    % }
    ; \textbf{Incorrect Cal} (CWE-682) (12\%), e.g., the potential DoS caused by incorrect calculations~\cite{2015NVDCVE-2015-8895}
    % \footnote{
    % % \url{https://github.com/imagemagick/imagemagick/commit/a321eb7fb3b35eed694d0366543450d023900b5b}
    % \url{https://nvd.nist.gov/vuln/detail/CVE-2015-8895}
    % }
    ; and \textbf{Control Flow} (CWE-691) (4\%), e.g., infinite loop caused by the incorrect data type of loop variable~\cite{2017NVDCVE-2017-12997}.
    % \footnote{
    % % \url{https://github.com/the-tcpdump-group/tcpdump/commit/fc6f112b40b9da56286ebc125f109e2889f5ebdb}
    % \url{https://nvd.nist.gov/vuln/detail/CVE-2017-12997}
    % } 
    \item \textbf{CodeChecker} warns most VCCs related to \textbf{Cond Check} (CWE-703) (20\%), e.g., null pointer dereferences caused by the incorrect order of arguments~\cite{2018NVDCVE-2018-7485}
    % \footnote{
    % % \url{https://github.com/lurcher/unixodbc/commit/4f9f77fb4204659ec9b7be8745d9e05a539c80b9}
    % \url{https://nvd.nist.gov/vuln/detail/CVE-2018-7485}
    % } 
    and \textbf{Coding Std} (CWE-710), (16\%) e.g., the improper  \lstinline{if} statement~\cite{2017NVDCVE-2017-13040}.
    % \footnote{
    % % \url{https://github.com/the-tcpdump-group/tcpdump/commit/3432f69892d872f2f46110774abc8cc676d9a533}
    % \url{https://nvd.nist.gov/vuln/detail/CVE-2017-13040}
    % }
\end{itemize}

\begin{table}[t!]
\centering
\footnotesize
\caption{Numbers of VCCs that receive warnings in all vulnerable functions (S8:\textit{allFn-Same}) or at least one vulnerable function (S6:\textit{1Fn-Same}) and percentages by CWE pillars. 
The first and the second numbers are from S8 and S6, respectively.
% Note that 13 VCCs related to Protection Mechanism Failure (CWE-693) do not receive warnings from any tools.
}
\begin{tabularx}{\linewidth}{
            >{\hsize=1.4\hsize}X
            >{\hsize=0.92\hsize}X
            >{\hsize=0.92\hsize}X
            >{\hsize=0.92\hsize}X
            >{\hsize=0.92\hsize}X
            >{\hsize=0.92\hsize}X
          }
\toprule
% \textbf{CWE Pillar} & \textbf{CCH} & \textbf{CQL} & \textbf{CPC} & \textbf{FLF} & \textbf{IFR}\\
\textbf{CWE Pillar} & 
\rotatebox{12}{\textbf{CodeChecker}} & 
\rotatebox{12}{\textbf{CodeQL}} &
\rotatebox{12}{\textbf{Cppcheck}} & 
\rotatebox{12}{\textbf{Flawfinder}} & 
\rotatebox{12}{\textbf{Infer}}\\
 \midrule
 \textbf{Access Ctrl}\newline
 CWE-284 & - & \textbf{1/1 \newline (3\%/3\%)}  & - & - & - \\ 
 \hline
 \textbf{Resource Ctrl}\newline
 CWE-664 & 17/44 \newline (3\%/4\%) & 43/63 \newline (8\%/12\%) & 4/8 \newline (1\%/1\%) &  \textbf{205/249 \newline (38\%/46\%)} & 24/33 \newline (4\%/6\%) \\ 
 \hline
 \textbf{Incorrect Cal}\newline
 CWE-682 & - & \textbf{ 9/11 \newline (12\%/14\%)} & 1/1 \newline (1\%/1\%) & 2/2 \newline (2\%/2\%) & - \\ 
 \hline
 \textbf{Control Flow}\newline
 CWE-691 & - & \textbf{2/2 \newline (4\%/4\%)} & - & - & - \\ 
 \hline
 % \textbf{CWE-693} & & & & & \\ 
 \textbf{Cond Check}\newline
 CWE-703 & \textbf{4/4 \newline (20\%/20\%)} & - & - & - & - \\ 
 \hline
 \textbf{Neutralization}\newline
 CWE-707 & 3/3 \newline (5\%/5\%) & - & - & \textbf{9/12 \newline (16\%/21\%) }& - \\  
 \hline
 \textbf{Coding Std}\newline
 CWE-710 & \textbf{8/12 \newline (16\%/27\%)} & 2/2 \newline(4\%/4\%) & 1/1 \newline (2\%/2\%) & - & 6/7 \newline (13\%/16\%) \\  
\bottomrule
\end{tabularx}
\newline
\label{table:detected-vcc-vulnerability-type}
\end{table}
% \vspace{-4mm}

\textbox{\textbf{Finding (vulnerability types)}: 
% All tools can detect vulnerabilty type \textbf{Resource Ctrl} (CWE-664). 
CodeQL can detect the most types of vulnerability in VCCs (5 out of 8). 
Only CodeQL can detect vulnerability type \textbf{Access Ctrl} (CWE-284) and \textbf{Control Flow} (CWE-691), while only CodeChecker can detect vulnerability type \textbf{Cond Check} (CWE-703). 
Flawfinder can detect most VCCs of type \textbf{Resource Ctrl} (CWE-664) and \textbf{Neutralization} (CWE-707).}

\inlineheading{Undetected vulnerabilities:} 
156 VCCs (22\%) spread across 33  projects (35\%) did not receive any warnings at the vulnerable functions (S5:\textit{1Fn-Any}). Notably, vulnerabilities related to CWE-664, CWE-682, CWE-693, CWE-707, and CWE-710 are the common types (143 out of the 156 VCCs) that were not detected by the studied \SAT{s}, which account for 10\% of the studied vulnerabilities.
% ,  were not detected by the studied \SAT{s}, which account for 10\%.
Surprisingly, some of these undetected vulnerabilities are straightforward and should have been identified by \SAT{s}. For instance, the CVE-2018-17249~\cite{2018NVDCVE-2018-17294} involves a buffer over-read due to a missing length check in a loop, deviating from the regular coding pattern~\cite{CTR51-CPP.Confluence}. 
Additionally, vulnerabilities related to \textbf{Protect Mech} (CWE-693) went unnoticed by \SAT{s}, despite the existence of supporting rules. Manual observation revealed that the current rules are strict to particular code patterns of vulnerabilities, a small deviation can make the vulnerability go undetected.
% did not cover \textcolor{red}{real-world} code changes introducing these vulnerabilities. 
As an example, CVE-2017-13083~\cite{2017NVDCVE-2017-13083} allows the execution of external code through an insecure connection. Existing CodeQL queries failed to detect these vulnerabilities, as the queries used regular expressions or focused on OpenSSL functions, while the vulnerable code checked HTTP certificates with Windows API.

\textbox{\textbf{Finding (undetected vulnerability)}: 22\% of VCCs do not receive warnings in the vulnerable functions. 
% 10\% of vulnerabilities are not detected. 
CWE-664, CWE-682, CWE-693, CWE-707, and CWE-710 are the common vulnerability types that the tools miss despite the presence of corresponding rules in the \SAT{s}. 
% VCCs of type \textbf{Resource Ctrl} (CWE-664) are missed most frequently, followed by \textbf{Incorrect Cal} (CWE-682) and \textbf{Neutralization} (CWE-707).  VCCs related to \textbf{Protect Mech} (CWE-693) do not receive any warnings although some \SAT{s} should be able to detect them.
}

\subsection{\RQTwoTopic{} (RQ2)} 
\label{rq2_result}

\smallheading{Approach}
\begin{figure}[t]
\center
  \includegraphics[width=0.9\linewidth,trim=0 10 0 10,clip=true]{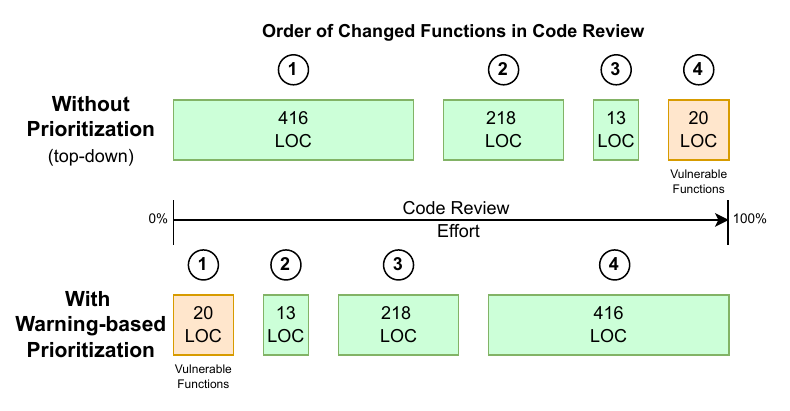}
\caption{An illustration to depict changed functions without prioritization (above) and with warning-based prioritization (below) using a real VCC from \textit{libsndfile}.}
\label{fig:changed-function-review-sequence}
\end{figure}
To answer our RQ2: \textit{\RQTwo}, 
% we measured code review performance for identifying vulnerable functions in VCCs when considering warnings of \SAT{s}.
we analyze how many vulnerable functions can be identified within a fixed review effort ($K$ lines of code; LOC) when the changed functions in a VCC are prioritized based on warnings of a \SAT{}.
% \SAT{} warning can help prioritize the changes that should be reviewed with a limited code review effort.
% For example, Figure~\ref{fig:changed-function-review-sequence} shows the order of changed functions with and without warning-based prioritization in a VCC.\footnote{\url{https://github.com/libsndfile/libsndfile/commit/1fc6bb8e0faa130e107f6af98991ac8e4555310b}} 
Figure~\ref{fig:changed-function-review-sequence} illustrates that the warning-based prioritization 
% with Flawfinder warning's density 
can affect the order of changed functions in code review of a VCC.\footnote{\url{https://github.com/libsndfile/libsndfile/commit/1fc6bb8e0faa130e107f6af98991ac8e4555310b}}
Without a prioritization (reviewing changed files alphabetically~\cite{AboutDocs}, top-down based on line number in each file), reviewers may stop reviewing the changes before reaching the vulnerable function or a vulnerable function may be reviewed the last, which requires more review effort to find a vulnerability.
In contrast, when a vulnerable function is prioritized by \SAT{} warnings, it may save code review effort in identifying vulnerabilities.
\reviewersModification{
We use the warnings from scenario \textit{S5 (1Fn-Any)} to answer this question because warnings may guide reviewers in prioritizing the code changes although the types differ from the reported vulnerabilities.}

% We hypothesize that \SAT{} warnings can help prioritize the changes that should be reviewed with a limited code review effort.
% Figure~\ref{fig:changed-function-review-sequence} shows the VCC\footnote{\url{https://github.com/libsndfile/libsndfile/commit/1fc6bb8e0faa130e107f6af98991ac8e4555310b}} that vulnerable function is the last function of a file. 
% The vulnerable changes may not be reviewed because the reviewer may stop reviewing during the first three functions.
% However, if the tool produces warnings in the vulnerable functions, the reviewer may review it before the other functions.
We explore three prioritization approaches based on the warning information~\cite{Trautsch2023AreProjects, Vassallo2020HowContexts}:
% that reviewers can potentially follow
\textcircled{1} \emph{Warning Amount (WA)}---a changed function with a high number of warnings reviewed first, \textcircled{2} \emph{Warning Density (WD)}---a changed function with a high number of warnings per LOC reviewed first~\cite{Trautsch2023AreProjects}, and \textcircled{3}  \emph{Warning Severity (WS)}---a changed function with high severity warnings\footnote{
To facilitate the analysis, we convert the severity levels of warning from each tool to the three levels i.e., High, Medium, and Low. For example, CodeQL's warning severity types \textit{Error}, \textit{Warning}, and \textit{Note} are mapped to \textit{High}, \textit{Medium}, and \textit{Low} respectively. The mapping table is included in the data package.
} reviewed first~\cite{Vassallo2020HowContexts}.
% (developers tend to investigate high severity warnings first~\cite{Vassallo2020HowContexts}). 

Inspired by prior studies~\cite{Wattanakriengkrai2022PredictingTechnique, Hong2022WhereTo}, we use the following metrics to measure the effectiveness of warning-based prioritization:

\begin{itemize}[noitemsep,topsep=1mm]
   \item \textbf{Initial False Alarm (IFA):} The proportion of lines of code in the changed functions that the reviewer needs to review until reviewing the first vulnerable function. 
   A high IFA indicates that more effort is needed to review non-vulnerable changes. 
    \item \textbf{Precision@$K$\%LOC:} The proportion of vulnerable functions that are prioritized within the top $K$\% LOC compared to the total number of changed functions prioritized within the top$K$\%LOC, i.e., $\tfrac{TP}{TP + FP}$.
    % where $TP$ is the number of vulnerable functions in top K\%LOC and $(TP+FP)$ is the total number of changed functions within the top K\%LOC. 
    % It can be calculated as $\tfrac{TP}{TP + FP}$ where TP is the number of vulnerable changes in top K\% and FP is the number of changes in top K\% that are not marked as vulnerable.
    \item \textbf{Recall@$K$\%LOC:} The proportion of vulnerable functions that are prioritized within the top $K$\% LOC compared to the total number of vulnerable functions in a VCC, i.e., $\tfrac{TP}{TP + FN}$. 
    % It can be calculated as $\tfrac{TP}{TP + FN}$ where TP is the number of vulnerable changes found in top K\% and FN is the number of vulnerable changes that are not in top K \%.
 
\end{itemize}

We set $K$=25\% LOC for Precision@$K$\%LOC and Recall@$K$\%LOC because the IFA of reviewing changed functions without warning-based prioritization is 25\% LOC on average \reviewersMinorModification{in our VCC dataset}, i.e., reviewers typically need to review 25\% of LOC of changed functions until the first vulnerable function is reviewed.
\reviewersMinorModification{
We also experiment with $K$=30\%, 40\%, and 50\% LOC to ensure the consistency.
}
% MOVE TO EXPLAIN EXAMPLE ABOVE
% A high IFA indicates that more effort is needed to review non-vulnerable changed functions. With high IFA, reviewers may stop reviewing before the vulnerable changes are reviewed.

\begin{figure}[t]
\center
  \includegraphics[width=\linewidth]{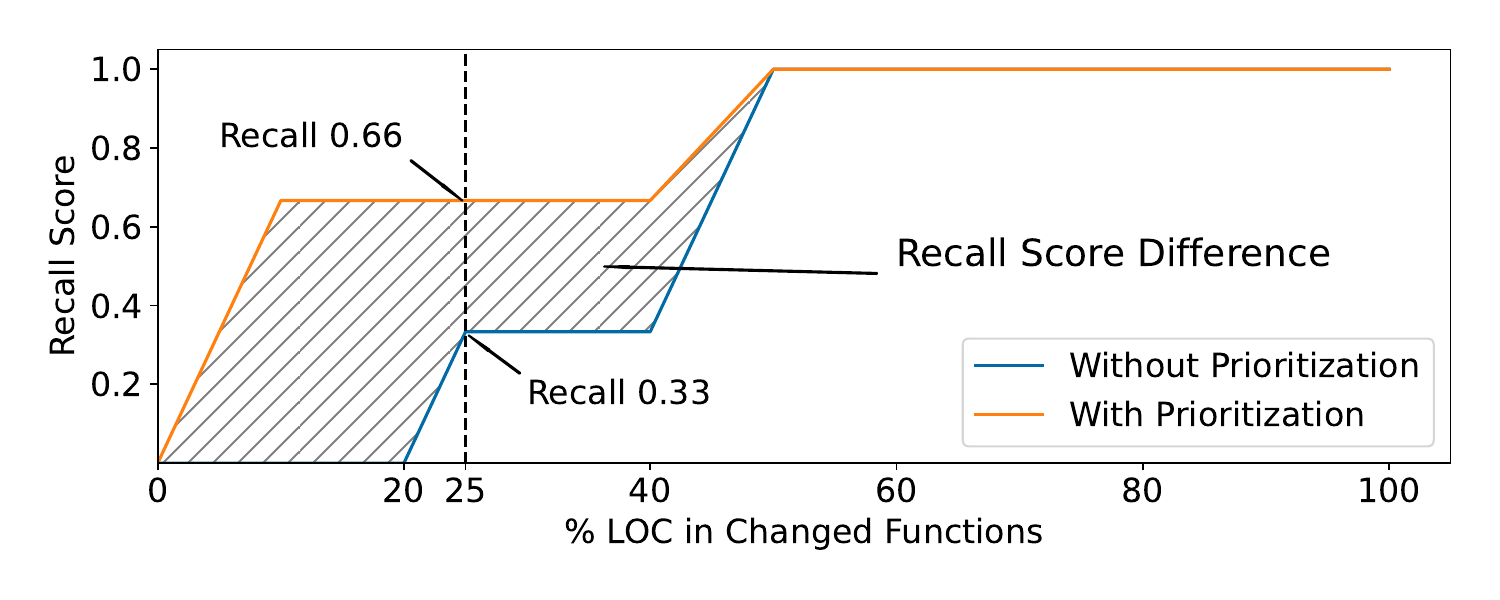}
\caption{An example of Recall@$K$\%LOC with and without warning-based prioritization of a real VCC from \textit{Leptonica}. 
}
\label{fig:performance-difference-example}
\end{figure}

To quantify the improvement of a warning-based prioritization by an \SAT{}, compared to non-prioritization, we measure the performance gain $= \frac{\Sigma_{i \in VCC} (Perf^i_{with} - Perf^i_{without})}{\Sigma_{i \in VCC} Perf^i_{without}}$\cite{Wattanakriengkrai2022PredictingTechnique}. Figure~\ref{fig:performance-difference-example} shows an example from a VCC\footnote{\url{https://github.com/danbloomberg/leptonica/commit/cef50b2cb3a8be397d81d4d32388a3918374e1b5}} when changed functions are prioritized by the severity levels of CodeChecker's warnings.
For example, at $K$=25\%LOC, the warning-based prioritization can improve the recall by 100\% ($\frac{0.66-0.33}{0.33}$), compared to a non-prioritization.
% It can be seen that more vulnerable functions can be reviewed when changed functions are prioritized with warning information at a given limited effort e.g., over 60\% higher recall at 20\% code review effort.

%  MOVE TO EXPLAIN fig:changed-function-review-sequence
% We hypothesize that warning information from the \SAT{s} can help prioritize the changes that should be reviewed with a limited code review effort.
% Figure~\ref{fig:changed-function-review-sequence} shows the VCC\footnote{\url{https://github.com/libsndfile/libsndfile/commit/1fc6bb8e0faa130e107f6af98991ac8e4555310b}} that vulnerable function is the last function of a file. 
% The vulnerable changes may not be reviewed because the reviewer may stop reviewing during the first three functions.
% However, if the tool produces warnings in the vulnerable functions, the reviewer may review it before the other functions.

% To check whether the performance of code review supported by tools at a fixed effort is improved from the code review without tool, we calculate the difference of performance metrics (i.e., Precision@K, Recall@K, and Initial False Alarm) between the prioritization approaches in the last section and the prioritization approach without warning information in each VCC. 

In addition, we use the Wilcoxon signed-rank test to confirm whether the performance differences between code reviews with and without tool support across VCCs are statistically different.
Since we compare the performance difference between the five tools, we used the Bonferroni correction to calculate the new acceptance threshold at a 95\% confidence level, as $\alpha = (\tfrac{0.05}{5}) = 0.01$.

\smallheading{Results}
% The percentages of performance difference between tools are shown in Table~\ref{table:percent-performance-difference}.
Table~\ref{table:percent-performance-difference} shows that IFA can be reduced by 13\% when the functions are prioritized based on CodeQL warnings.
% We also observe that the IFA of reviewing changed functions without warnings-based prioritization is 25\% LOC on average, indicating that the reviewers typically need to review 25\% of LOC of changed functions until the first vulnerable function is reviewed.
% Thus, we set $K$=25\% LOC for Precision@K\%LOC and Recall@K\%LOC.
When the reviewing effort is fixed at 25\% LOC of changed functions, warning-based prioritization can increase precision by 12\% when using CodeQL warnings. 
Recall scores can also be increased by 5.6\% when using CodeChecker warnings.
Table~\ref{table:percent-performance-difference} also shows that the precision and recall of warning-based prioritization is statistically higher than non-prioritization.
Nevertheless, we also observe that the performance of some warning-based prioritization (e.g., based on the warning amount of Flawfinder) can be lower than non-prioritization.
% In particular, ranking changed functions by warnings from some tools (e.g., Flawfinder, CodeQL, and Infer) with certain ranking approaches may reduce code review recall scores.
We also check the performance with other $K$ values (i.e., $K$ = 30, 40, and 50\% LOC) and the statistical analyses confirm that the performance of warning-based prioritization is better than non-prioritization.

\begin{table*}[ht!]
    \centering
    \footnotesize
    \caption{Percentage of performance differences when changed functions are ranked with warning-based prioritization in scenario S5 (\textit{1Fn-Any}) at 25\% of LOC in changed functions. }
    \begin{tabularx}{\linewidth}{
            >{\hsize=1\hsize}X
            >{\hsize=1\hsize}X
            >{\hsize=1\hsize}X
            >{\hsize=1\hsize}X
            >{\hsize=1\hsize}X
            >{\hsize=1\hsize}X
            >{\hsize=1\hsize}X |
            >{\hsize=1\hsize}X
            >{\hsize=1\hsize}X
            >{\hsize=1\hsize}X
          }
          \toprule
         \textbf{Tool}
        &  \textbf{Prec@25\% (WA) $\nearrow$}
        &  \textbf{Prec@25\% (WD) $\nearrow$}
        &  \textbf{Prec@25\% (WS) $\nearrow$}
        &  \textbf{Recall@25\% (WA) $\nearrow$}
        &  \textbf{Recall@25\% (WD) $\nearrow$}
        &  \textbf{Recall@25\% (WS) $\nearrow$}
        &  \textbf{IFA \newline (WA) $\searrow$}
        & \textbf{IFA \newline (WD)$\searrow$}
        & \textbf{IFA \newline  (WS)$\searrow$}
\\
\midrule
\textbf{CodeChecker}& 9.36\%\ddag &	8.09\%\ddag	&	7.87\%\ddag	&	\goodNumbers{\textbf{2.65\%}}	&	\goodNumbers{\textbf{5.66\%}}	&	\goodNumbers{\textbf{2.00\%	}}&	\goodNumbers{\textbf{-6.87\%}} &	-10.29\%\dag	&	\goodNumbers{\textbf{-5.81\%}}	\\
\textbf{CodeQL}& \goodNumbers{\textbf{12.09\%\ddag}}	&	\goodNumbers{\textbf{10.98\%\ddag}}	&	\goodNumbers{\textbf{9.93\%\ddag}}	&	-4.72\%	&	3.43\%	&	-3.37\%	&	-2.51\%	&	\goodNumbers{\textbf{-13.34\%\ddag}}	&	-2.36\%	\\
\textbf{Cppcheck}& \badNumbers{2.10\%}	&	\badNumbers{2.10\%}	&	\badNumbers{2.10\%}	&	0.81\%&	0.81\%	&	0.81\%	&	-0.44\%	&	\badNumbers{-0.39\%}	&	-0.44\%	\\
\textbf{Flawfinder}& 9.87\%\dag	&	2.51\%	&	8.84\%	&	\badNumbers{-8.58\%}	&	\badNumbers{-0.38\%}	&	\badNumbers{-7.94\%}	&	\badNumbers{7.85\%}	&	-4.23\% &	\badNumbers{6.21\%}	\\
\textbf{Infer}& 9.04\%\ddag	&	9.14\%\ddag	&	9.05\%\ddag	&	-1.98\%	&	3.25\%	&	-1.87\%	&	-3.24\%	&	-10.79\%	&	-3.36\%	\\
\bottomrule
 \multicolumn{10}{l}{
 \ddag~strongly significant: $p$ < 0.002; \dag~moderately significant: 0.002 $\le$ $p$ < 0.01 ($\alpha$ is adjusted with Bonferroni correction)
 } \\
    \end{tabularx}
    \label{table:percent-performance-difference}
\end{table*}

% Mini analysis: Number of Warnings
% Regarding the number of warnings, which is a major concern for \SAT{} users~\cite{Johnson2013WhyBugs, Chess2004StaticSecurity, Panichella2015WouldReviews}, we find that some tools can produce more warnings than others. 
% Specifically, Flawfinder produces a larger number of warnings per function (median=3) in more changed functions (median=4) of a VCC.
% Hence, extra effort is required to inspect the warnings that may be irrelevant to the vulnerable changes.

% \textbox{\reviewersModification{\textbf{Finding (potential drawback)}: Despite detecting most VCCs, Flawfinder produces more warnings in more functions, which may negatively impact warning-based prioritization.}
% }

\textbox{\textbf{Finding (warning-based prioritization)}: Compared to non-prioritization, warning-based prioritization using CodeQL warnings yields a substantial improvement for precision (12\%) and Initial False Alarm (13\%). CodeChecker offers a substantial improvement for recall (5.6\%).

}

% \begin{figure}[h]
% \center
%   \includegraphics[width=\linewidth]{figure/vcc_warning_number_scenarios.pdf}
% \caption{Average number of warnings and warned changes in VCCs}
% \label{fig:execution-time-commit-size}
% \end{figure}

The different prioritization approaches can yield different performance gains.
As seen in Table~\ref{table:percent-performance-difference}, prioritizing changed functions within VCCs with \textit{Warning Density (WD)} usually yields the largest improvement for recall (up to 5.6\%) and IFA (up to 13.3\%) regardless of \SAT{s}.
In terms of precision, \textit{WD} provides a slightly lower improvement than \textit{Warning Amount (WA)} but \textit{WA} normally offers a lower recall score and a higher IFA.
On the other hand, \textit{Warning Severity (WS)}, which developers typically use during the development process~\cite{Vassallo2020HowContexts}, yields the lowest performance improvement \reviewersMinorModification{compared to other prioritization approaches}
across every metric.

\textbox{\textbf{Finding (prioritization approach)}: Prioritizing changed functions using \textit{Warning Density (WD)} generally yields the largest improvement among the three approaches. 
% should help reviewers find the vulnerable functions faster.
}

\subsection{\RQThreeTopic{} (RQ3)} 
\smallheading{Approach}
\reviewersModification{
To answer our RQ3: \textit{\RQThree{}}, we measure the computation time of each experiment i.e., executing a tool on one VCC.
This period encompasses the entire process including the other necessary steps.
}

\smallheading{Results}
\reviewersModification{
% The descriptive statistics of computation time are shown in Table~\ref{table:tool-execution-time-statistics}.
% Flawfinder is the fastest tool, with an average computing time of 20 seconds, attributed to its syntactic analysis of subjects without the need for compilation. 
% Conversely, CodeChecker, CodeQL, and Infer, which require the compilation process, share comparable average computation times ranging from 213 to 456 seconds. 
% Counterintuitively, Cppcheck, which employs both semantic and syntactic techniques without compiling the subjects, turns out to be the slowest tool, with an average computation time of 2,702 seconds.
% % Friedman test indicates that the execution times of studied tools on the selected VCCs are significantly different. 
% We use the Wilcoxon signed-rank test to statically compare the computation time of each \SAT{} with each of the other tools on each VCC.
% % the fastest tool on each VCC.
% We find that Flawfinder's computation time is statistically lower than other tools, followed by CodeChecker. 
% CodeQL's and Infer's computation times are not significantly different. 
% On average, Flawfinder is 135 times faster than the slowest tool, followed by CodeChecker (13 times) while CodeQL ties with Infer (8 and 6 times, respectively). 
% % Cppcheck usually takes the longest time to execute.
Table 7 presents the computation time statistics. 
Flawfinder took the least computation time, with an average of 20 seconds, due to its syntactic analysis approach that bypasses compilation.
On the other hand, CodeChecker, CodeQL, and Infer which require compilation took comparable average computation times ranging from 213 to 456 seconds.
Surprisingly, Cppcheck, employing both semantic and syntactic techniques without compilation, took the longest computation time, with an average of 2,702 seconds.
% To assess the significance of these differences, we employ the Wilcoxon signed-rank test to compare the computation time of each \SAT{} with every other tool on each VCC.
% Flawfinder's computation time is significantly lower than other tools, with CodeChecker following. CodeQL and Infer show no significant difference in computation times. 
% On average, Flawfinder is 135 times faster than the slowest tool, followed by CodeChecker at 13 times, while CodeQL and Infer tie at 8 and 6 times faster, respectively.
}

\begin{table}[h]
\centering
\footnotesize
\caption{Computation time statistics}
\begin{tabularx}{\linewidth}{
            >{\hsize=1.4\hsize}X
            >{\hsize=0.92\hsize}X
            >{\hsize=0.92\hsize}X
            >{\hsize=0.92\hsize}X
            >{\hsize=0.92\hsize}X
            >{\hsize=0.92\hsize}X
          }
\toprule
% \textbf{Time (s)} & \textbf{CCH} & \textbf{CQL} & \textbf{CPC} & \textbf{FLF} & \textbf{IFR}\\
\textbf{Time (s)} &
\rotatebox{12}{\textbf{CodeChecker}} & 
\rotatebox{12}{\textbf{CodeQL}} &
\rotatebox{12}{\textbf{Cppcheck}} & 
\rotatebox{12}{\textbf{Flawfinder}} & 
\rotatebox{12}{\textbf{Infer}}\\
 \midrule
 \textbf{Miminum} &   3  &  26  & 1 &  \textbf{1} &  2 \\ 
 \cline{2-6}
 \textbf{Mean} & 213 & 343 & 2,702  &  \textbf{20} & 456 \\
 \cline{2-6}
 \textbf{Median} &  74    &  148    & 1,546 & \textbf{6} & 184 \\ 
 \cline{2-6}
 \textbf{Maximum} & 7,268  & 3,314 & 16,974  & \textbf{88}  & 2,305  \\
\bottomrule
\end{tabularx}
\newline
\label{table:tool-execution-time-statistics}
\end{table}

\textbox{\textbf{Finding (average time)}: The average C/C++ \SAT{} computation time is between 20 seconds and 45 minutes. Flawfinder is the fastest tool.}

% A recent study by \citet{Kudrjavets2022MiningAnalysis} reported that the average \textit{time-to-first-response} in open-source projects is between 42 and 125 hours.
% Hence, it can be implied that in most cases \SAT{s} can be integrated into the CI/CD pipeline without affecting the waiting time of reviewers.

% \textbox{\textbf{Finding (waiting window)}: \SAT{} computing time fits within the average time-to-first-response in code reviews.}

\begin{figure}[h]
\center
  \includegraphics[width=\linewidth,trim=0 10 20 30, clip=true]{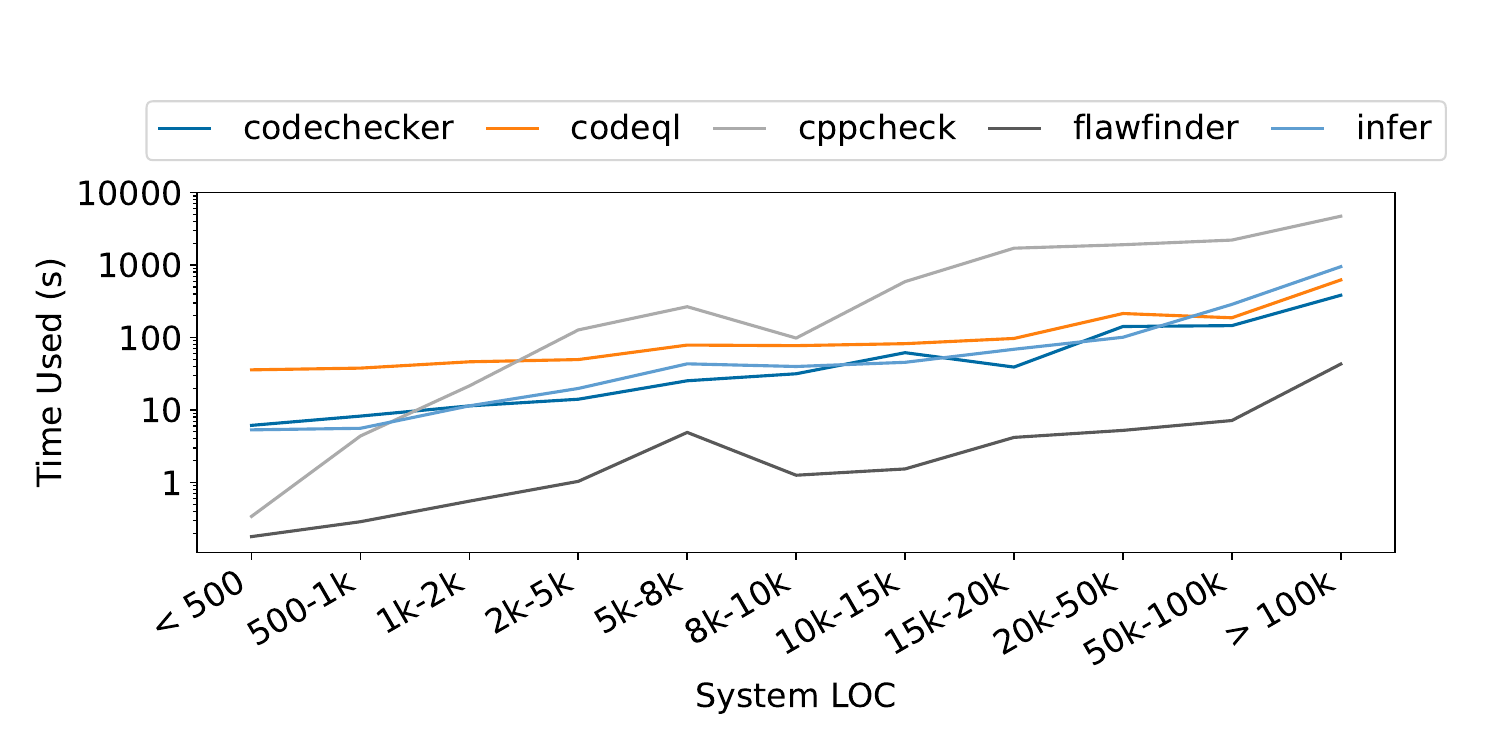}
\caption{An average computation time across various system sizes in VCCs (Total LOC)}
\label{fig:execution-time-commit-size}
\end{figure}

\reviewersModification{
% Computation time may be varied by the size of systems.
% Therefore we measure the size of the system in each VCC and examine the trend of computation time across the corresponding VCCs.
% To understand how the system's size influences \SAT{}'s computation time, we determine the average computation time of each tool on systems of different sizes.
% As shown in Figure~\ref{fig:execution-time-commit-size}, Cppcheck is visibly slower (i.e., average computation time is more than 10 minutes) when systems are larger than 10k-15k LOC while CodeQL, Infer, and Codechecker are slower for systems that are larger than 50k-100k LOC.
% In any case, Flawfinder's computation time is seemingly constant.
% The long computation time, especially when the size is larger than 10k-15k LOC for Cppchek may be because it analyzes files in sequence,\footnote{\url{https://sourceforge.net/p/cppcheck/discussion/general/thread/13e5f86e88/\#73f9 }} hence the large projects may have more files that incur the bottleneck and influence the total computation time.
Computation time varies with system size. 
We measure system size (lines-of-code; LOC) in each VCC and examine computation times across VCCs. 
The average computation time on systems of different sizes reveals that Cppcheck is notably slower (over 10 minutes) for systems larger than 10k-15k LOC, while CodeQL, Infer, and Codechecker are slower for systems larger than 50k-100k LOC. 
In contrast, Flawfinder's computation time remains relatively constant. 
The Cppcheck's prolonged computation time, especially for systems exceeding 10k-15k LOC, may be due to its sequential file analysis approach, resulting in bottlenecks with larger projects.
}

% It is counterintuitive because while Flawfinder---a syntactic-based tool---is the fastest Cppcheck which is a hybrid tool is slower than other semantic-based tools in general.

\textbox{\textbf{Finding (system size)}: The tools require longer computation time when the size of the system is larger than 50kLOC-100kLOC.}

% We seek to explain the possible cause of the noticeably long computation time in Cppcheck, especially with projects that are larger than 10k-15k LOC.
% It was found that the developer of Cppcheck had explained\footnote{\url{https://sourceforge.net/p/cppcheck/discussion/general/thread/13e5f86e88/\#73f9 }} that Cppcheck analyzes files in sequence.
% Therefore, a bottleneck can occur on specific files that require a longer time to analyze.
% The projects with a larger LOC may have more files that incur the bottleneck and influence the total computation time.

\section{Discussion}
\label{discussion}
\subsection{\SAT{} Performance for Code Review}
\label{tool_performance_discussion}

\reviewersModification{
Based on our empirical results, each \SAT{} has shown promises and limitations for code review.
Although Flawfinder has the highest detection rate (Section~\ref{fig:vcc-detection-scenarios}) and takes the least computation time, warning-based prioritization using Flawfinder yields relatively low performance (Section~\ref{table:percent-performance-difference}). 
This is partly because Flawfinder produces a larger number of warnings to
% ($median=3$) 
the changed functions
% ($median=4$) 
than the other \SAT{s}.
% Despite Flawfinder's high detection rate reported in RQ1, the large number of warnings in changed functions may lead to its low performance in warning-based prioritization.
% This suggests that the warnings of Flawfinder may not be the best option for prioritizing the changed functions under a limited review effort.
% Consequently, it can be inferred that the abundance of warnings generated by Flawfinder may generally deter reviewers from utilizing \SAT{s}~\cite{Christakis2016WhatStudy, Johnson2013WhyBugs, Chess2004StaticSecurity, Panichella2015WouldReviews}.
This suggests that Flawfinder's abundance of warnings may not be ideal for prioritizing changed functions under limited review effort. 
% and could discourage reviewers~\citep{Christakis2016WhatStudy, Johnson2013WhyBugs, Chess2004StaticSecurity, Panichella2015WouldReviews} from using the tool.
On the other hand, while CodeQL and CodeChecker have lower detection rates compared to Flawfinder, their warnings effectively aid in prioritizing changed functions.
% CodeQL achieves the second-highest detection rate in all scenarios disregarding the warning types.
% CodeQL can produce warnings in most CWE pillars (Table~\ref{table:detected-vcc-vulnerability-type}).
% It also has the highest precision and lowest Initial False Alarm (IFA) in warning-based prioritization 
When using CodeQL warnings for prioritization, the Initial False Alarm (IFA) and precision of finding vulnerable functions were also improved compared to non-prioritization.
Similarly, CodeChecker can 
% produce warnings in the second-most CWE pillars and 
% has the highest recall performance gain in warning-based prioritization.
offer a substantial improvement in recall for warning-based prioritization.
Our findings suggest that \SAT{s} could assist secure code reviews by helping prioritize changed functions based on warnings.
However, further refinement is necessary to better support these practices.
% These results suggest that each tool offers different benefits and emphasize that the high detection rates at the function level may not indicate better support for code reviews.
}

\reviewersModification{
Although we analyze the warnings and vulnerability types based on CWE pillars in this study, the CWE item of a warning within the same CWE pillar as the VCC may not match the CWE item of the vulnerability in the corresponding CVE~\citep{Li2023ComparisonJava}.
% Thus, we examine \textbf{the accuracy between the CWE items of the warning and vulnerability}.
Thus, we examine whether the CWE item of the warning matches the CWE item of the vulnerability to understand \textbf{the accuracy of CWE pillar proxy}.
% the \textbf{accuracy of CWE pillars} with the actual CWE items of VCCs and the \SAT{} warnings. 
We sample 400 warnings from the 7,564 warnings that all \SAT{} produced in vulnerable functions regardless of vulnerability types (\textit{S5:1Fn-Any}), which allow us to draw a conclusion with a 95\% confidence level and a 5\% margin of error~\cite{DiBiase2016AChromium}.
Following \citet{Li2023ComparisonJava}'s approach, we determine if the CWE item of a warning matches the CWE item of a VCC using the CWE item descriptions.
% We analyze whether the CWE items of the warnings and the CWE items of the VCCs differ. 
% Following \citet{Li2023ComparisonJava}'s approach, a warning will be labeled as a match when the CWE item of the warning is related to the CWE item of the VCC based on the descriptions.  
% NOTE OMIT examples to save space - already explained in the response letter
% For example, \texttt{Improper Limitation of a Pathname to a Restricted Directory ('Path Traversal')} (CWE-22) is related to \texttt{Path Traversal: '..filename'} (CWE-29).
% Although the CWE pillars of the warning and VCC differ, it is also possible that the CWE items of the warning and VCC match.
% For example, \texttt{Improper Neutralization} (CWE-707) is related to \texttt{Use of Externally-Controlled Format String} (CWE-134).
We find that 66.42\% of the warnings within the same CWE pillars as the VCC match the CWE items of the VCCs.
We also find that 10.27\% of the warnings with the different CWE pillars from the VCC actually match the CWE items of the VCCs.\footnote{The evaluation results are included in the data package.}
% However, the warnings are more accurate (\textcolor{red}{aa%-\textcolor{red}{cc%) in identifying vulnerabilities within the same CWE pillars. 
}

\reviewersModification{
To provide more rigorous insights beyond the detection rate at the function level, we further investigate \textbf{the relevancy of the warnings} on the lines of code in VCCs (i.e., the warned lines) to the vulnerability reported in the CVEs.
To do so, we manually analyze whether the warnings actually reflect the vulnerabilities in the VCCs. 
We examine the following information: the warned lines, the description of vulnerability in the associated CVE, and the fixing commit.
We classify the warnings into three groups (following~\citet{Thung2012ToTools}): 1) \textbf{Relevant}--the warned line and the associated warning are relevant to the vulnerability and the fixing commit, 2) \textbf{Irrelevant}--the warned line and the warning are completely irrelevant to the vulnerability, and 3) \textbf{Unsure}--the warning is relevant to the vulnerability, but the warned line is not; and vice versa.
The relevant group estimates a true positive rate of the warnings. 
The irrelevant group estimates the false positive rates of the warnings. 
We report the unsure group as it may still be beneficial in code reviews where the reviewers gain more security awareness from the adjacent warnings~\citep{Braz2022SoftwarePerspective}.
Based on the same sample set of 400 warnings, we find that 
7\% of the warnings are relevant to the warned line and vulnerability,
76\% of the warnings are irrelevant to the warned line and vulnerability,
and 17\% of the warnings are either relevant to the warned line or vulnerability.
These results, along with the findings in RQ2 (Section~\ref{rq2_result}), suggest that warning information could aid reviewers in prioritizing which functions to review first. However, it remains essential for reviewers to invest effort in meticulously examining and identifying vulnerabilities within these prioritized functions.
}

\subsection{Implications \& Suggestions} 
\label{implication_suggestion}
We discuss the implications for practitioners, \SAT{} developers, and researchers in this section.

% DISCUSSABLE POINTS
% it is expected that the required manual effort for code review activity should be reduced~\cite{Singh2017EvaluatingEffort, Balachandran2013ReducingRecommendation}.

% Several studies~\cite{Panichella2015WouldReviews, Singh2017EvaluatingEffort, Johnson2013WhyBugs, Chess2004StaticSecurity, Imtiaz2019HowUsage} noticed that \SAT{s} can produce a large amount of irrelevant results.

% different tools are suitable for certain types of issues and require different efforts for practical use~\cite{Antunes2015AssessingExamples, Balachandran2013ReducingRecommendation, Liang2018Fuzzing:Art, Lipp2022AnDetection}.

\subsubsection{For practitioners}
% Our results show that \SAT{s} can effectively detect vulnerabilities in code commits (RQ1), improve code review performance (RQ2), and require relatively short computation time (RQ3).
Our results highlight the potential of using \SAT{s} for secure code reviews.
To maximize the benefits of \SAT{}-supported secure code reviews, we provide the following recommendations.
% However, the unique constraints of code review context i.e., limited effort and time of reviewers must be considered.  
% POINT 1 - START FROM EFFORT \& PRIORITIZATION EFFECTIVENESS
\textcircled{1} \textbf{\SAT{s} can be leveraged to reduce code review effort.} 
Table \ref{table:percent-performance-difference} shows that at a fixed effort (25\% LOC of changed functions), warning-based prioritization can increase the efficiency in identifying vulnerable functions compared to non-prioritization. 
RQ2 shows that prioritizing based on warning density generally provides a substantial improvement, suggesting that this prioritization approach could be adopted.  
% Reviewers can prioritize vulnerable functions based on the warning density in the changed functions.
Additionally, RQ3 shows that \SAT{}'s computing time is relatively short which should fit within the code review waiting period~\cite{Kudrjavets2022MiningAnalysis}.
\textcircled{2} \textbf{Changed functions that receive \SAT{s}' warnings should be carefully investigated regardless of the warning types. } 
RQ1 shows that warning types may not be directly relevant to the vulnerability of the VCCs.
% during the development process because a vulnerability can be a result of multiple code commits. 
Figure~\ref{fig:vcc-detection-scenarios} shows that regardless of the warning types, \SAT{s} can produce warnings in the vulnerable functions (S5:\textit{1Fn-Any}), however, only 1\%-46\% of the VCCs receive warnings with the same type as their CVEs (S6:\textit{1Fn-Same}).
% 44\% of VCCs with warnings in vulnerable functions (S5: \textit{1Fn-Any}) may not receive warnings of the correct type. 
% \reviewersModification{The warning relevancy evaluation also shows that 7\% of warnings in vulnerable functions are relevant to the vulnerability regardless of the warning types.}
This result suggests that reviewers should not limit the focus of security review to the warning types that the tool produced.
% Nevertheless, from our manual observations, the warnings with a different type from the CVE vulnerable functions that receive any warning types, the imprecise warning types still relate to the vulnerability in the VCCs. 
 % although not accurately pinpointing the expected type, we observed during a manual inspection that some warnings in this group can address the vulnerability.
 % COMBINED TWO SUGGESTIONS BELOW
\textcircled{3} \textbf{\SAT{} should be carefully selected based on the project's security needs and resource constraints.
}
We found that several factors can impact the \SAT{s}'s effectiveness, e.g., vulnerability types, available time, and computation resources.  
% Running \SAT{s} requires various types of resources such as computing power and waiting time.
Software projects should determine their security needs and resource constraints when choosing \SAT{s}.
% Table~\ref{table:detected-vcc-vulnerability-type} shows that each SAT can correctly detect different types of vulnerability. 
% while CodeQL can detect more \textbf{Incorrect Cal} (CWE-682) vulnerabilities.
If the 
% correctness of vulnerability types is the main concern
\reviewersModification{matching vulnerability type is the main concern},
specific security needs, such as types of vulnerability that lead to critical impact, should be recognized. 
For example, Table~\ref{table:detected-vcc-vulnerability-type} shows that only CodeQL can detect vulnerability type \textbf{Access Ctrl} (CWE-284) and \textbf{Control Flow} (CWE-691), while only CodeChecker can detect vulnerability type \textbf{Cond Check} (CWE-703).  
Alternatively, when computing power is abundant, combining \SAT{s} can increase the vulnerability detection effectiveness by 26\% (Figure~\ref{fig:tool-vcc-venn}).
If the project is exceptionally large, the computation time should be deliberately assessed as RQ3 indicates a substantial increase for projects exceeding 50kLOC-100kLOC.
\reviewersModification{In addition, we find that setting up \SAT{s} may demand technical expertise. Nevertheless, with the support of build automation tools, the set up of \SAT{s} should be a one-time implementation effort.}

\subsubsection{For \SAT{}'s developers}
Despite the potential values of \SAT{s}, we find that incorporating them in secure code reviews still needs extra support.
% To improve \SAT{s}, 
We offer the following recommendations to \SAT{}'s developers.
% \textcircled{1} \textbf{\SAT{s} should allow users to easily define the rules that serve their needs.}
% RQ1 shows that \SAT{} may miss unprecedented vulnerabilities; especially, project-specific implementations. 
% Therefore, \SAT{} should support user-defined rules.
% % \SAT{} should identify vulnerable code changes although they do not follow coding patterns.
% % Personalizing rules to the projects may also help.
% Although some \SAT{s} let users create new rules, prior work~\cite{Beller2016AnalyzingSoftware} reported that most projects only rely on pre-defined rules, partly because creating new rules is difficult.
% % Hence, the rule's creation procedure should be simplified.
% % as project-specific implementations may not align with the pre-defined rules. 
% Thus, the process of creating new rules should be simplified. 
% % However, such procedures are still complicated for typical users.
% % \textcolor{red}{TRY USING EXISTING RESULTS AS BOLD TEXT e.g., Fig 3 / MAYBE IMPROVE COMMUNICATION / CONTEXT }
\textcircled{1} \textbf{Provide vulnerability advice in warnings.} 
\reviewersMinorModification{
Our manual evaluation shows that some warnings are directly relevant to VCCs.
However, some warnings with unmatching CWE pillars can also identify the vulnerabilities.}
Figure~\ref{fig:vcc-detection-scenarios} also shows that the VCCs with correct types of vulnerability can be reduced up to 35\% at the function level, i.e., CodeQL in S5 (\textit{1Fn-Any}) and S6 (\textit{1Fn-Same}).
Even if the relevant issue is detected, imprecise warning information can distract reviewers.
\SAT{} should highlight the risk if an issue can lead to vulnerability, otherwise the warning can be perceived as non-vulnerable.
% although they signify the issues that lead to vulnerability. 
\textcircled{2}\textbf{Warning severity information should be improved.}
Table~\ref{table:percent-performance-difference} shows that prioritizing changed functions with warning severity does not yield a substantial improvement compared to other approaches.
Since developers usually follow high-severity warnings during the development process~\cite{Vassallo2020HowContexts}, warning severity should be more reliable.

% \SAT{} users may not be able to select the tools that suit their needs because of the lack of vulnerability coverage information.

% simple technique vs advanced technique

% \textbf{Undetected issues}
% - What types of issues are not detected?
% - Why SAT cannot identify some code changes that contribute to vulnerability?
% Based on the results (RQ1), vulnerabilities related to protection mechanisms and improper exception handling are more likely missed by the tools. 

% - because these vulnerabilities are based on the specific implementation of each project, making it more difficult to create the checkers?
% - more checkers are needed to cover real-world code changes
% - custom checkers development~\footnote{\url{https://github.blog/2021-11-16-adding-custom-codeql-queries-code-scanning/}}

% - Coverage VS Overlapping between tools -> make suggestions from Venn Diagram
% - Come up with some assumptions -> e.g., two tools with similar techniques [vulnerability types, if possible - but keep it simple] can find more .. etc.

% \textbf{Improve ease of use}
% - Tool setup difficulty, reflection from the study
% - Default settings, performance trade-off, customizable options - can users create new checkers?
% - Integrability - what tools can be integrated with popular code review platforms

\subsubsection{For researchers}
Some practices in \SAT{}-supported secure code reviews are not well understood. 
We suggest that future work explore the following aspects.
\textcircled{1} \textbf{Rules for unconventional/unfinished code should be explored.}
% Brace for unconventional code.} 
Code changes can be untidy during the development cycle and code reviews. 
In our RQ1, we observe that developmental code that deviates from pre-defined rules can be a cause of undetected vulnerabilities.
\reviewersModification{
AI models (e.g., Large Language Models) may help in detecting unprecedented vulnerabilities beyond the developer-defined rules of \SAT{s}.
However, a recent work~\citep{Wang2023TheAnalysis} suggests that their vulnerability detection capability might be restricted by an insufficient semantic understanding of complex code.}
Therefore, future studies should further investigate the solutions that detect vulnerabilities in such kind of code.
% MERGED WITH NEXT SUGGESTION
% \textcircled{2} \textbf{How reviewers use \SAT{s} to identify vulnerable changes.}
% Our investigation is based on the literature reporting that reviewers prioritize changed functions using \SAT{} warnings. 
% However, it is unclear what is the best way to use such warnings. 
% Future work should investigate this practice to understand other benefits of \SAT{s} in code reviews.
% This knowledge may guide \SAT{} developers to provide better support for seccode reviews.
% \textcircled{3} \textbf{Other techniques to enhance \SAT{s} for code reviews.} xxx
\textcircled{2} \textbf{Effective techniques to use \SAT{s} for secure code review should be investigated.} 
Table~\ref{table:percent-performance-difference} shows that warning-based prioritization with the three approaches can improve precision, recall, and IFA compared to non-prioritization.
Meanwhile, certain approaches may cause performance drawbacks e.g., prioritizing changed functions by warning amount of Flawfinder can reduce recall and increase IFA.
\reviewersModification{
Our manual evaluation also highlights the high false positive rates of the warnings (over 76\%) in vulnerable changes which can strongly hinder the effectiveness of code reviews~\citep{Christakis2016WhatStudy}.
These results and the findings in RQ2 (Section~\ref{rq2_result}) suggest that warnings can assist reviewers in prioritizing functions, but reviewers must carefully examine and identify vulnerabilities within prioritized functions.}
Indeed, effective techniques to use \SAT{s} for secure code review are still largely unexplored, which is an open challenge for future work.
% As this study is the first to report these results, it is an open challenge to improve these performance gains, either with other information in \SAT{} warnings or other techniques.
\reviewersMinorModification{
\textcircled{3} \textbf{Undetected vulnerabilities should be prioritized.} 
Our result emphasizes that all \SAT{s} can still miss many vulnerabilities 
% are still \textit{unsound}, i.e., there exist false negative results~\citep{Barr2015TheSurvey, Shahriar2010ClassificationDetectors}, 
in code changes by showing that 22\% of VCCs do not receive the warnings in the vulnerable code changes from any tools (Section~\ref{rq1_result}).
% Our manual evaluation of the warnings also shows that they are \textit{incomplete} i.e., there exist false positive warnings~\citep{Barr2015TheSurvey, Shahriar2010ClassificationDetectors}.
% Alternate solution for vulnerability identification such as [--TODO complete or sound solutions --] may prevail to support the practitioners.
% [--Discuss Soudiness proposal - Livshits--]
% \cite{Livshits2015InSoundiness}
The failure to detect real-world vulnerabilities stresses the importance of improving \SAT{s}.
Future work should explore the solutions to these shortcomings and minimize the undetected vulnerabilities.
}
\textcircled{4} \textbf{An ensemble approach of multiple \SAT{s} should be investigated.} 
Figure~\ref{fig:tool-vcc-venn} shows that combining tools can improve the effectiveness of vulnerability identification.
However, it is infeasible for projects with limited computing resources to use many \SAT{s}, especially for large projects (Figure~\ref{fig:execution-time-commit-size}).
A practical guideline to combine \SAT{s} is useful for practitioners.

\section{Threats to Validity} 
\label{threats}
We discuss potential threats to this study's validity in this section.

\inlineheading{Internal validity:}
The VCC dataset quality impacts study results. 
We verified vulnerable changes, discarded VCCs that could not be linked to fixing commits, and 
excluded a redundant commit caused by a renamed project. 
For \SAT{} selection, we carefully chose the popular \SAT{s} that practitioners can employ in the development process.
We were aware that enabling different \SAT{}'s rules can affect the effectiveness in general. 
We adhered to default rules because it is strongly recommended by tool developers. 
In addition, when attempting to enable extra rules, we encountered out-of-memory issues.
\reviewersMinorModification{
Upon manual mapping of the warnings to the CWEs pillar (Section~\ref{warning_type_mapping}), human biases can potentially occur. 
To address this, the two authors first lay the groundwork for mapping before proceeding collaboratively with the remaining warnings.}
\reviewersModification{Similarly, we acknowledge the potential error during the manual evaluation of the warnings.}
Lastly, as CWE and CVE entries are regularly updated by MITRE, result sustainability cannot be guaranteed.
% because our analyses used the available data at the time of this study. 

% The quality of the VCC dataset impacts the results of this study. 
% To ensure the correctness of VCCs, we verified the vulnerable change locations and discarded the VCCs that could not be linked to the fixing commits. 
% We noticed one redundant commit in our selected VCCs, caused by a renamed project.
% We excluded this redundant commit from our analyses.

% Indeed, different sets of \SAT{s} may lead to different results. 
% We carefully selected popular \SAT{s} that practitioners could have used in the development process.
% We were also aware that enabling different \SAT{}'s rules can affect the effectiveness in general. 
% However, we opted to use the set of rules that each tool readily provides because it is strongly recommended by the tools' developers.
% It should be noted that we encountered out-of-memory issues in an attempt to enable extra rules.

% Finally, the CWE and CVE entries are regularly updated by MITRE.
% Our analyses used the available data at the time of this study. 
% Hence, we cannot guarantee that the results will be sustained.

\inlineheading{Construct validity:}
The construct validity concerns the measured indicators.
% \textcolor{red}{Due to the limited insights into how reviewers use \SAT{} warnings}
Based on the previous literature~\cite{Muske2016SurveyAlarms, Trautsch2023AreProjects, Vassallo2020HowContexts}
, we assume that reviewers prioritize the changed functions with the warnings during code reviews.
We analyzed the prioritization performance at various fixed efforts ($K$\% LOC) to ensure consistency.
To increase the validity, we explore various prioritization approaches, i.e., warning amount, density, and severity.
However, reviewers may adopt other approaches when reviewing code changes, introducing potential variations beyond the examined constructs.
% warning-based prioritization may not be a technique in which the reviewers utilize \SAT{} warnings or our prioritization approaches may not represent the real-world scenarios.
    
\inlineheading{External validity:}
The external validity concerns the generalizability of the results. 
Our findings are supported by the breadth of our dataset which contains 815 commits from 92 diverse projects that contributed to 319 vulnerabilities.
\reviewersModification{
Our results are based on five \SAT{s}.
A broader set of \SAT{s} may improve the generalizability of the findings.
Nonetheless, these five \SAT{s} are commonly used, actively maintained, and compatible with the studied systems and executable via CLI~(Section~\ref{tool_selection_criteria}).
To support future work, we release a framework~\cite{Charoenwet2024AnCommits} that future work can extend for additional \SAT{s} and code change datasets in diverse languages.
We also acknowledge that the findings regarding the relevancy of the warning (Section~\ref{tool_performance_discussion}) are based on the sample warnings, which may have limited generalizability.}
% We only present the findings as an approximation of the warnings' relevancy to the VCCs.
For transparency, and to facilitate future work, we release the benchmark, datasets,
\reviewersModification{
and evaluation results.
}

\section{Related Work} 
\label{related_work}

The effectiveness of \SAT{s} on VCCs has not been previously investigated.
\reviewersModification{
Most existing works primarily focused on evaluating the \SAT{s} on the released versions of software or the crafted programs that contain known vulnerabilities.
The datasets of code commits submitted by developers for code reviews are not usually used by \SAT{} studies.
}
\citet{Lipp2022AnDetection} evaluates five free and one commercial C/C++ \SAT{s} and reported that \SAT{s} can miss 47\%-80\% of vulnerabilities. 
Their benchmark dataset includes 111 front-ported vulnerabilities from Magma dataset~\cite{Hazimeh2020Magma:Benchmark} and 81 vulnerabilities from released versions of Binutils and FFmpeg.
In this work, we used the dataset of vulnerability contributing commits which include vulnerable functions and vulnerable files, and found that \SAT{s} can detect at least one vulnerable function of 78\% of VCCs.
\citet{Kannavara2012SecuringAnalysis} evaluated the effectiveness of \textit{Klocwork}, a multi-language \SAT{}, on the Linux Kernel.
% to investigate vulnerability detection early in the software development process. 
However, only one version of the Linux Kernel was used.
In this work, we investigate multiple versions of code (815 VCCs) from the development process.
% We argue that the vulnerable versions of programs alone may not represent the code changes, which are the subject of code reviews.
% , that gradually contribute to the vulnerability.
A recent study by \citet{Li2023ComparisonJava} investigated a closely relevant context to our work. 
They evaluated seven \SAT{s} using code commits and reported that \SAT{s} detect only 12.7\% of the vulnerabilities with the correct vulnerability types. 
% which is significantly lower than the results from the synthetic programs. 
However, they focused on Java \SAT{s} and seven types of CWE vulnerability.
In this work, we study C/C++ \SAT{s} using the VCC dataset with eight types of CWE vulnerability.

Differing from previous works, 
\reviewersModification{we focus on real-world developmental code commits that are the subject of code reviews. This novel dataset can enhance the understanding of \SAT{}'s performance, offering insight into \SAT{}'s performance in the development phase of a software project.}
% we study C and C++ \SAT{s} with code commits that contributed to exploitable vulnerabilities in eight types of CWEs.
% We find that prioritizing the changed functions in a commit with \SAT{} warning information can improve accuracy (i.e., higher precision and recall and lower Initial False Alarm) compared to prioritizing changed functions without \SAT{} warnings. 
% The precision and Initial False Alarm are statistically different.
We are also the first to investigate warning-based prioritization and show that the effectiveness in identifying vulnerabilities can be improved when the changed functions are ranked with the information from \SAT{} warnings.
% In terms of vulnerability identification effectiveness by a single tool, our finding partly aligns with a finding from \citet{Lipp2022AnDetection}.
% We find that Flawfinder can detect more VCCs than other tools, while Cppcheck can identify the least VCCs regardless of vulnerability types or granularity levels.
% This finding partly aligns with a finding from \citet{Lipp2022AnDetection} who reported that Flawfinder can detect the vulnerabilities with correct types at the function level more than other tools and that Cppcheck can detect the least vulnerabilities.
% Additionally, we also report the time performance of the C and C++ \SAT{s} that has not been reported previously. 

\section{Conclusion}
\label{conclusion}
In this study, we conducted an empirical study to understand the practical benefits of C and C++ \SAT{s} for secure code reviews.
% Our dataset contains \textcolor{red}{815} vulnerability-contributing commits that contributed to \textcolor{red}{319} exploitable CVE vulnerabilities in \textcolor{red}{92} open-source projects.
The results show that 52\% of VCCs can be warned by a single tool in the changed functions that contain vulnerable code.
By combining tools, the detection effectiveness can increase by 26\%.
For secure code reviews, prioritizing changed functions with \SAT{} warnings can improve accuracy (i.e., 12\% of precision and 5.6\% of recall) and reduce Initial False Alarm (i.e., lines of code in non-vulnerable functions that must be inspected until the first vulnerable function) by 13\%.
Moreover, the average computing time of \SAT{s} should fit within the waiting time of the code reviews.
\reviewersModification{However, at least 76\% of the warnings in vulnerable functions can be irrelevant to the vulnerability in VCCs, and 22\% of the VCCs can be undetected.}

% Based on these findings, we provide suggestions for practitioners and \SAT{} developers, including the researchers who conduct future works on \SAT{}-supported secure code reviews.

% \textcolor{red}{Based on these findings, we suggest that reviewers consider \SAT{} warnings when examining code changes, as they can help reviewers effectively spot vulnerabilities in a limited effort. 
% Combining multiple tools should enhance overall detection. 
% However, reviewers must be aware that developmental code or project-specific code may not align with pre-defined \SAT{} rules.
% % To address this, \SAT{} developers should simplify the rule creation process.
% Lastly, future research can explore the other practical benefits of SATs for secure code reviews. }
% REWRITTEN FROM SUGGESTED POINTS
Based on these findings, we suggest software projects select \SAT{s} that fit their security needs and resource constraints, leverage \SAT{s} to streamline code review efforts, and inspect changed functions with \SAT{} warnings, irrespective of warning types.
On the other hand, \SAT{}'s developers should enhance vulnerability advice in warnings and improve severity information.
Future work can \reviewersModification{improve \SAT{s} on the undetected vulnerabilities}, explore rules for developmental code, investigate effective techniques for using \SAT{s} in secure code reviews, and explore an ensemble approach involving multiple \SAT{s}.
We also release the \SAT{} evaluation framework~\cite{Charoenwet2024AnCommits} and dataset to support forthcoming investigations.

\section*{Data Availability}\label{data_availability}
We have released the datasets and scripts used in this study to foster future works on \SAT{}-supported secure code reviews~\cite{Charoenwet2024DatasetReview}.

\section*{Acknowledgment}\label{acknowledgment}
This work was supported by the use of Nectar Research Cloud, a collaborative research platform supported by NCRIS-funded Australian Research Data Commons (ARDC).
Patanamon Thongtanunam and Van-Thuan Pham were supported by two Australian Research Council’s Discovery Early Career Researcher Award (DECRA) projects (DE210101091 and DE230100473). 
We sincerely thank anonymous reviewers for their valuable feedback.

\onecolumn \begin{multicols}{2}

% \bibliography{references.bib}
\bibliography{references}
\bibliographystyle{ACM-Reference-Format}

\end{multicols}

\end{document}